\documentclass{iopjournal}

\usepackage{ragged2e}

\usepackage[center, labelsep=endash]{caption} 
\usepackage[list=true]{subcaption}

\usepackage{booktabs}
\usepackage{longtable} 
\usepackage{array}

\usepackage{amssymb}
\usepackage{siunitx}
\usepackage{verbatim}

\begin{document}

\articletype{Article} 

\title{Two-length spatial correlation function of turbulence in TCV}

\author{Olivier Panico$^{1,*}$\orcid{0000-0000-0000-0000}, Pascale Hennequin$^2$\orcid{0000-0000-0000-0000} Sascha Rienäcker$^{2}$\orcid{0000-0000-0000-0000}, Oleg Krutkin$^{1}$, Benoit Labit$^{1}$, Yanick Sarazin$^{3}$ and the TCV team$^{4}$}

\affil{$^1$Ecole Polytechnique Fédérale de Lausanne (EPFL), Swiss Plasma Center (SPC), Lausanne, CH-1015, Switzerland}

\affil{$^2$Laboratoire de Physique des Plasmas (LPP), CNRS, Sorbonne Université, Ecole polytechnique, Institut Polytechnique de Paris, Palaiseau, France}

\affil{$^3$CEA, IRFM, F-13108 Saint-Paul-lez-Durance, France}

\affil{$^4$See author list of B.P. Duval et al 2024 Nucl. Fusion 64 112023}

\affil{$^*$Author to whom any correspondence should be addressed.}

\email{olivier.panico@epfl.ch}


\begin{abstract}
\begin{justify}
Spatial correlation functions of density fluctuations are measured in the Tokamak à Configuration Variable (TCV) using a dual-channel Doppler backscattering (DBS) diagnostic. In certain cases, the spatial correlation function exhibits two characteristic length scales. By analogy with nonlinear reduced simulations, the presence of two correlation lengths may be indicative of avalanche-like transport. The correlation functions obtained from DBS are compared with those from short-pulse reflectometry measurements and show reasonable agreement. Both short- and long-range correlations are measured in the same plasma geometry for different heating powers. Short-scale correlation lengths are found to be on the order of $3$--$5$ Larmor radii, while large-scale correlations extend over approximately $5$--$15$ Larmor radii. The correlations are found to decrease towards the very edge of electron cyclotron heated discharges, coinciding with a narrow $E_r$ well. 
\end{justify}
\end{abstract}

\tableofcontents


\section{Introduction}

Turbulence and its associated transport are among the most critical phenomena that limit the performance of magnetic confinement fusion devices. For tokamaks in low-confinement mode (L-mode) two fundamental mechanisms can be identified. The first involves small-scale mixing-type transport governed by the size of turbulent eddies and the fluctuations' amplitude. This is generally considered local: a steeper local pressure gradient drives faster instability growth rate and consequently larger transport. The second mechanism operates at large scales through an avalanching mechanism \cite{bak1987self, carreras1996model, hahm2018mesoscopic}. In this picture, a local flattening of the pressure profile generates two flanking regions of steeper gradients. Each region drives a larger transport, which in turn flattens the adjacent profile. As a result, a pressure bump propagates down the gradient and a pressure void propagates in the opposite direction \cite{diamond1995dynamics}. The overall process gives rise to near-ballistic transport, which significantly increases confinement losses. \\

This theoretical view has been extensively studied in simulations. To quantitatively reproduce avalanche transport, flux-driven simulations are required -- where gradients are free to evolve rather than being externally fixed -- as well as a global, non-local framework \cite{sarazin1998intermittent}. Several gyrokinetic codes have demonstrated avalanche transport \cite{idomura2009study,mcmillan2009avalanchelike,  jolliet2012plasma, ku2009full, gorler2011flux, dif2010validity, di2024system}. Additionally, reduced modelling can be used to capture part of the avalanche physics. Among these, one can note early cellular automata models such as in ref.\cite{newman1996dynamics}, reduced interchange \cite{sarazin2000transport} and now popular flux-driven Hasegawa-Wakatani fluid models such as described in refs.\cite{qi2020dimits, ghendrih2022role, panico2025importance, guillon2026anisotropic} \\


Experimentally diagnosing these two mechanisms presents a significant challenge. For small-scale turbulence, the primary quantities of interest are the characteristic size of turbulent structures, the associated level of density and temperature fluctuations and their cross-phase. In the core and edge regions, temperature fluctuations can be obtained using electron cyclotron emission (ECE) \cite{PhysRevLett.33.758, freethy2019advances} while reflectometry is widely used for density fluctuations \cite{mazzucato1998microwave}. The size of the structures is often inferred from the correlation version of the two diagnostics in which two close locations are probed using either multiple diagnostic channels or a fast-sweep configuration \cite{hornung2013turbulence}. In particular, conventional reflectometry -- at normal incidence to the plasma cutoff surface -- was extensively used in the early 2000s to characterize the size of turbulent structures \cite{conway1999reflectometer}. However, this method is particularly challenging to interpret due to the nonlinear response of the reflectometer to the amplitude of density fluctuation and due to small angle forward scattering along the beam path \cite{nazikian1995reflectometer}. Doppler backscattering (DBS) addresses this issue by sending the beam at an oblique angle with respect to the cutoff surface and measuring the backscattered rather than the reflected signal, greatly reducing forward scattering contributions \cite{gusakov2002non, altukhov2016poloidal, krutkin2019nonlinear}. In addition, DBS is also inherently wavenumber selective as a given probing angle selects a perpendicular wavenumber $k_\perp$ of the turbulent spectrum \cite{conway2004plasma, hennequin2004doppler}. \\

Diagnosing large-scale avalanching transport is even more demanding. Avalanches are transient, radially-extended (and possibly poloidally and toroidally meandering) events, and capturing their propagation requires diagnostics with simultaneously high spatial and temporal resolution across a broad radial range. Therefore, it is essential to correlate signals across multiple radial locations. For these reasons, experimental evidence for avalanche-like transport is scarce. A major contribution was made by P.A.Politzer on the DIII-D tokamak, where heat avalanches were reported using the ECE diagnostic \cite{politzer2000observation}. \\ 

In this contribution, we report on correlation measurements performed on the Tokamak à Configuration Variable (TCV) \cite{theiler2026progress} using a dual channel DBS. The measurements are performed under both neutral beam heating (NBH) and electron cyclotron heating (ECH) to modify the temperature profiles while keeping a similar plasma geometry and density profile. It is found that in certain cases, a second slope appears on the spatial correlation function of the turbulence. This type of two-length correlation function is also observed in simulation results when the transport is dominated by avalanches. In \autoref{section: measuring correlation length}, the method used to measure correlation lengths with DBS is reviewed. In \autoref{section: comparison with}, the spatial correlation functions of DBS are compared with short-pulse reflectometry (SPR). In \autoref{section: scaling of}, the experimental conditions are described and the correlation lengths are compared at different heating scheme and power, together with the fluctuations perpendicular velocity. The correlation lengths are found to be reduced in the $E_r$ well, provided that it is sufficiently narrow. Finally, in \autoref{section: avalanches simulations}, similar techniques are applied to simulations in cases that are either dominated by avalanche-like transport or not. The former are found to exhibit a two-length spatial correlation function.

\section{Measuring correlation lengths using DBS}
\label{section: measuring correlation length}


\subsection{DBS correlation measurements on TCV}

The measurements are performed with a dual V-band channel Doppler backscattering diagnostic (DBS) \cite{hirsch2001doppler, conway2004plasma, hennequin2004doppler} that has recently been installed on the \textit{Tokamak à Configuration Variable} (TCV). The diagnostic principles and the specific setup on TCV are detailed in ref.\cite{rienacker2025survey}. DBS operates by launching a microwave beam into the plasma at an oblique angle with respect to the plasma cutoff surface. When the beam approaches the cutoff layer, it is backscattered by density fluctuations whose perpendicular wavenumber $k_\perp$ satisfies the Bragg condition: $k_\perp = -2k_i$ where $k_i$ is the incident probing wave-number at the beam turning point. The backscattered signal then provides a measure of the amplitude of density fluctuations at the selected $k_\perp$, and the velocity perpendicular to the magnetic field can be inferred from the Doppler shift of the received signal. By varying the launch angle, the probed perpendicular wavenumber can be selected in a typical range $k_\perp \in [2 - 15]\SI{}{cm^{-1}}$.

With the addition of a second DBS channel, in a configuration known as \textit{correlation DBS} (CDBS), the spatial correlation function of density fluctuations can be measured \cite{schirmer2006radial, schneider2021overview}. The radial correlation is obtained by simultaneously sending two microwave beams from the same line of sight using different frequencies. A fixed \textit{reference} channel probes a fixed plasma location, while the \textit{hopping} channel scans around the reference. \\

On TCV, the DBS is located on the upper low field side of the vessel and probes from the top. 
Given the antenna constraints in terms of accessible angles, the choice is made in this contribution to use an upper single null (USN) shape so that the DBS probes towards the midplane.

\begin{figure}[h]
	\centering
	\includegraphics[width=0.5\textwidth]{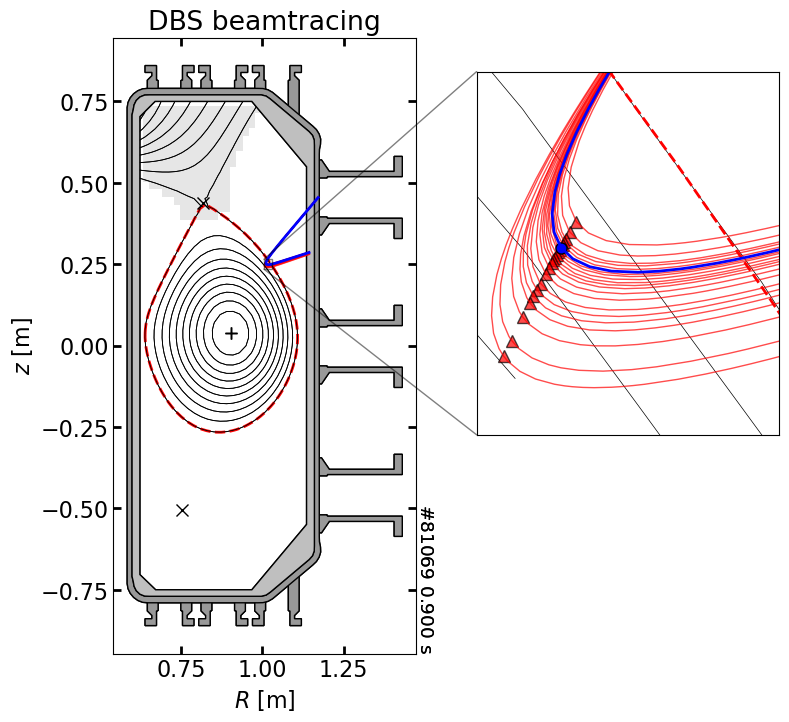}
	\caption{Beamtracing for $\#81069$, $t = [0.8 - 0.9]$s, poloidal angle $\theta = -50^{\circ}$. The blue ray indicates the reference channel, $F_{ref}=$\SI{60}{GHz}. The red rays correspond to the hopping channel whose frequency is sequentially stepped between \SI{59.05}{GHz} and \SI{63.05}{GHz}.}
	\label{fig: ex beamtracing correlation}
\end{figure}

Since the location of the cutoff layer depends both on the launched frequency and the local density profile, different frequencies probe different radial positions. In the present configuration, the diagnostic provides coverage of the plasma edge, $\rho \in [0.7 - 1]$ for $k_\perp \in [6 - 10]\SI{}{cm^{-1}}$. To determine the precise measurement location and the effective $k_\perp$, a beamtracing code \cite{honore2006quasi} is used, accounting for the beam geometry and plasma density profile as well as the reconstructed equilibrium magnetic field. An example of beamtracing is shown in \autoref{fig: ex beamtracing correlation} for shot $\#81069$.  The poloidal section of the diverted upper single null plasma is shown on the left, and a zoom on the turning points is indicated in the inset. Two channels are used; the blue one (circle) is the \textit{reference} and the red one (triangles) the \textit{hopping}, whose frequency is stepped to probe sequentially along the different red ray traces.

\begin{figure}[h]
	\centering
	\includegraphics[width=0.5\textwidth]{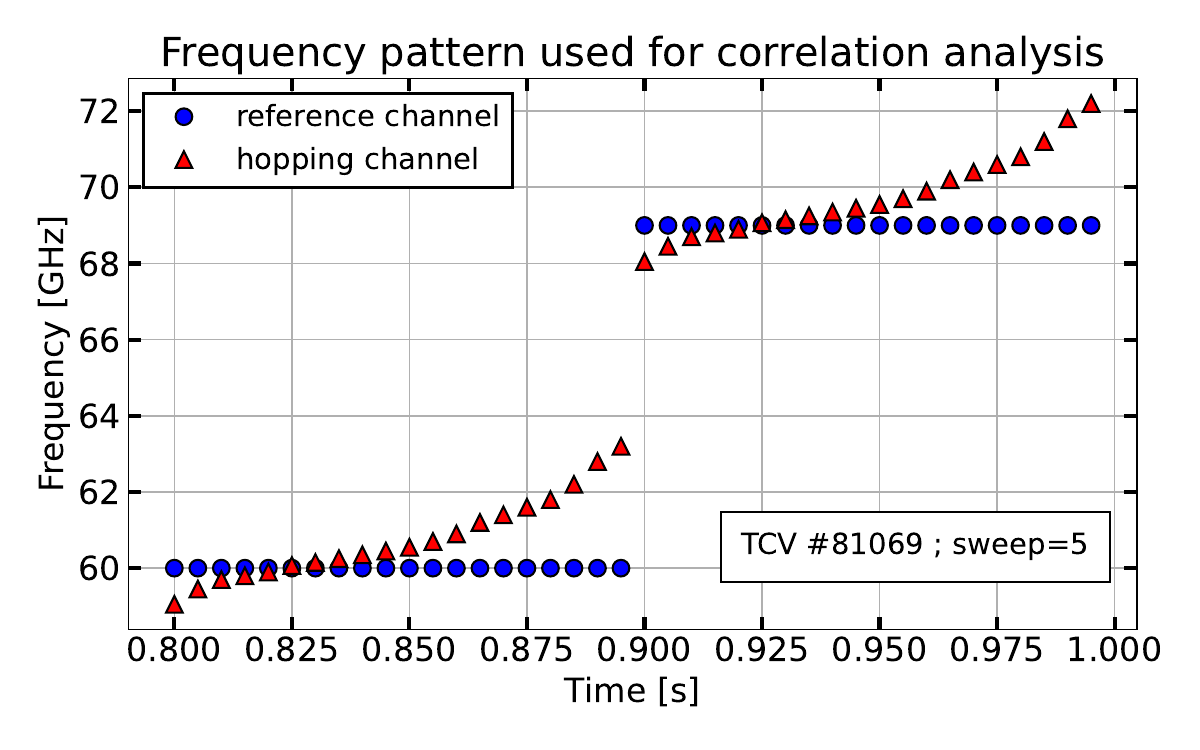}
	\caption{Frequency pattern used for correlation. Reference channel is indicated in blue (circles), hopping channel is in red (triangles). Shot: TCV $\#81069$, $t = [0.8 - 1]$s.}
	\label{fig: frequency pattern}
\end{figure}

The frequency pattern corresponding to the above beamtracing is provided in \autoref{fig: frequency pattern}. 

In this example, the plasma is probed sequentially at two reference locations: $F_{ref}=\SI{60}{GHz}$ and $F_{ref}=\SI{69}{GHz}$. The blue circles correspond to the reference channel at a fixed frequency, while the red triangles indicate the hopping channel. Estimating the spatial correlation function of the turbulence requires averaging over many turbulence decorrelation times to achieve statistical convergence. Each of the 20 probing frequencies acquires $\SI{5}{ms}$ of data at a sampling rate of $\SI{40}{ns}$, giving a total acquisition time of $\SI{100}{ms}$ per complete spatial correlation measurement. The frequency separation between the 20 points is chosen to ensure sufficient radial coverage of the measurement region. Throughout this interval, the plasma parameters and geometry must remain stationary. Most importantly, the density must remain stationary during the acquisition time so that the hopping channel can scan around the reference channel. Correlation functions where plasma parameters exhibit significant drifts over this timescale are excluded from the analysis. \\ 

\begin{figure}[h]
	\centering
	\includegraphics[width=0.8\textwidth]{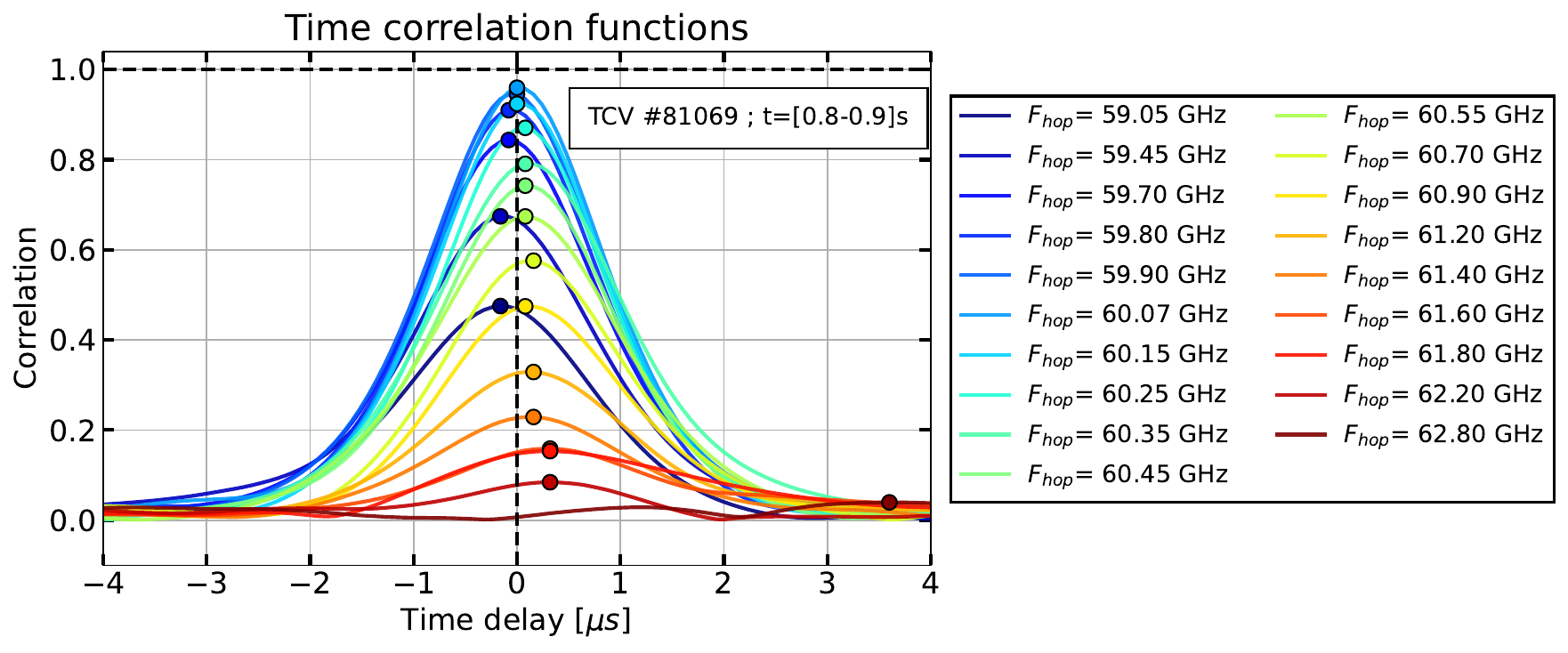}
	\caption{Complex signal correlation functions between reference and different hopping channel frequencies $F_{hop}$. The correlation maximum is indicated with a circle for each curve. The color code from blue to red is related to the hopping channel frequency, from small to large.}
	\label{fig: example time correlation functions}
\end{figure}

The method used to compute the correlation between the reference and the hopping channels is briefly reviewed here using the example shot $\#81069$, $t=[0.8-0.9]$s, poloidal angle $\theta = -50^{\circ}$, $F_{ref}= 60$ $GHz$. Each complex signal is split into segments of $n_{FFT}=1024$ points with an overlap of $512$ points. The power spectral density (PSD) of the reference and hopping channels are computed together with their cross-spectral density (CSD) for each segment that are then averaged out using Welch's method. The CSD is then inverse Fourier transformed to obtain the amplitude of the time-domain correlation function, shown in \autoref{fig: example time correlation functions}. The circles indicate the maximum of the correlation amplitude. 

In \autoref{fig: example time correlation functions}, the peak correlation decreases as the probing frequency separation increases. When the hopping frequency is lower than the reference frequency, the time lag at which the correlation reaches its maximum is negative; it becomes positive once the hopping frequency exceeds the reference value. This lag corresponding to the maximum correlation could be used to estimate the tilt angle of turbulent structures, as explained in refs.\cite{pinzon2019measurement, pinzon2019experimental}. Note that this view has been challenged in ref.\cite{krutkin2020theoretical}. In the present experiments, the DBS probes the plasma obliquely from the top of the vessel. In this configuration, the delay is very small making the tilt angle difficult to estimate. \\

The spatial correlation function is obtained from the maximum of the time correlation function as a function of the distance between the reference and hopping channel. This distance is estimated as a function of $R$ and $Z$:
\begin{align}
	\Delta = \sqrt{ (R_{ref} - R_{hop})^2 + (Z_{ref}-Z_{hop})^2} \; ,
\end{align}
with $R$ and $Z$ denoting the probed coordinates of the turning points for the reference and hopping channels in the poloidal plane, as obtained from the beamtracing code. 



\subsection{Use of the MUSIC algorithm to compute correlation at large scale}
\begin{figure}[h!]
	\centering
	\includegraphics[width=0.5\textwidth]{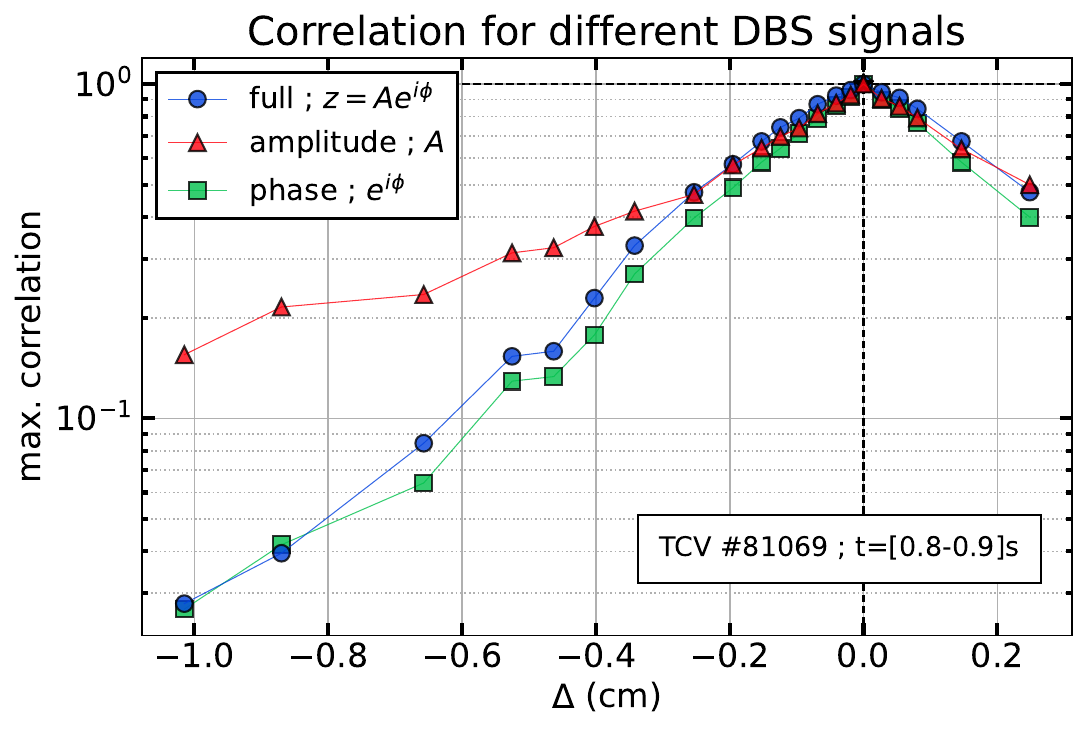}
	\caption{Spatial correlation function obtained from the max. of the correlation function. The computation is applied to the full complex signal (blue circles), the amplitude signal (red triangles) and the exponential of the phase (green squares).}
	\label{fig: radial correlation full amp phase}
\end{figure}

Several approaches have been used in the literature to estimate the spatial correlation length from DBS measurements. The heterodyne detection provides in-phase $x(t)$ and in-quadrature $y(t)$ signals which yields the complex $z(t) = x(t) + i y(t) = A(t) \exp(i \phi(t))$ from which either the amplitude $A(t)$ or phase $\phi(t)$ can be extracted. In Doppler reflectometry, the amplitude is sensitive to the density fluctuations intensity while the phase carries information on both the fluctuations and the velocity. The full complex signal $z$ has been used in refs.\cite{estrada2001turbulence, schirmer2007radial} to study the radial correlation length of the turbulence at short scales while ref.\cite{schneider2021overview} uses the signal amplitude. To assess the influence of this choice on the results, \autoref{fig: radial correlation full amp phase} compares the spatial correlation functions computed with the complex, amplitude and phase signals for the test case. 

In this example, all signals exhibit similar short-range correlations. However, the amplitude correlation maximum exhibits a second slope that decays more slowly than the complex and phase signals, which at large separation fall below the noise correlation level of $\sim 0.1$ (from a mixed files estimate). This different behaviour can be understood as follows. The complex signal carries information on both the amplitude of the fluctuations and their associated Doppler shift. At large radial separations, or in the presence of strong velocity shear, the Doppler shift may differ between the reference and hopping channels, shifting their respective PSD maxima to different frequencies. This frequency mismatch reduces the CSD amplitude and effectively masks long-range correlations. Working with the signal amplitude alone removes this Doppler contribution. The resulting correlation is therefore larger and provides a more reliable estimate of the radial extent of density perturbations. The amplitude signal is used in the remainder of the paper. Note that not all cases exhibit a two-slope correlation function (see \autoref{fig: example double slope}). \\

\begin{figure}[h!]
	\centering
	\includegraphics[width=0.5\textwidth]{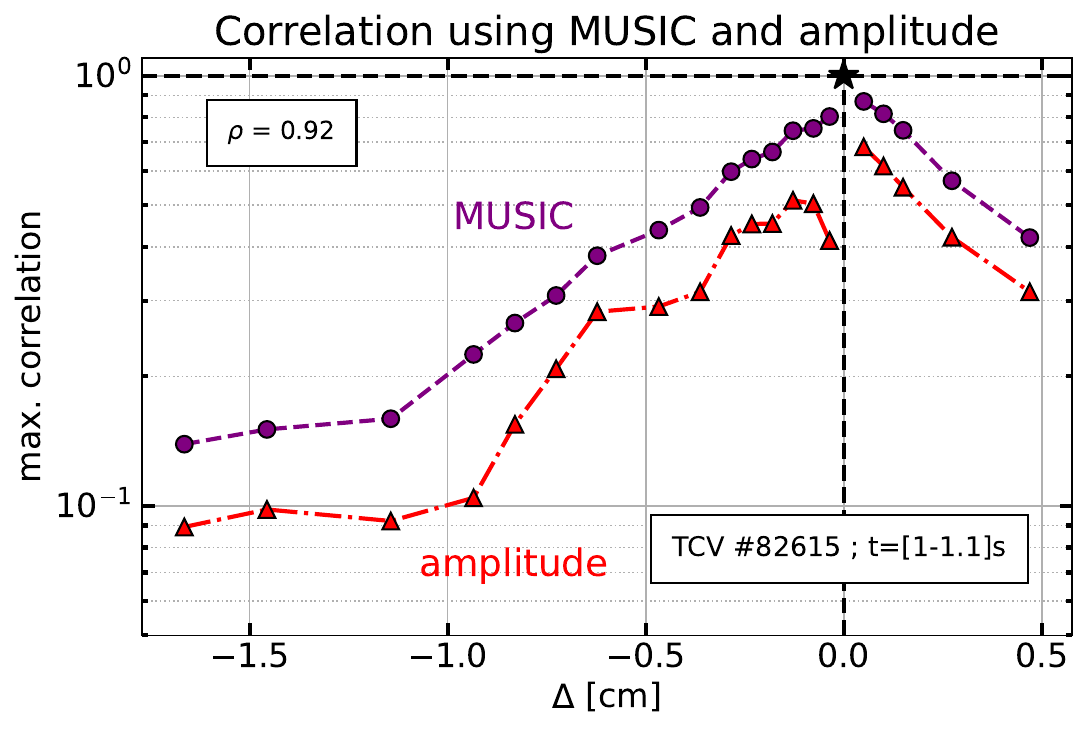}
	\caption{Spatial correlation functions computed with the raw amplitude signal or using the amplitude inferred from the MUSIC algorithm as a function of separation $\Delta$. Discharge $\#82615$ $t=[1-1.1]$s.}
	\label{fig: comparison music amp}
\end{figure}

When the signal to noise ratio becomes low, the correlation level is affected, and the maximum of the correlation function remains low even if the channels are very close. This is seen in \autoref{fig: comparison music amp} in red. This makes the evaluation of the correlation length difficult, since the slopes become artificially flat. To improve the evaluation of the correlation when the signals are low, typically when probing towards the core ($\rho < 0.9$), we can use signal processing methods which better extract the dynamics of the signals and their correlation. Here, we use the MUltiple SIgnal Classification (MUSIC) algorithm which was already introduced for the analysis of DBS data in ref.\cite{hennequin2006fluctuation} to compute the instantaneous Doppler velocity, in ref.\cite{vermare2012detection} to detect geodesic acoustic modes and in ref.\cite{panicothesis2024} to analyse long-range correlations on the instantaneous velocity. Details on the MUSIC algorithm and how it preserves the relevant part of the DBS signal are given in \autoref{appendix: details on the MUSIC algorithm}. \\

MUSIC is a spectral estimation method based on the eigen analysis of a data vector separating it into distinct signal and noise sub-spaces. The algorithm can provide a fine estimation of the signal frequency content especially for short time series and is particularly adapted to signals having a limited number of frequency components, such as speech or music, where the amplitude and frequency of the signal evolve on a longer timescale than the inverse of dominant frequency. The is also the case for DBS data whose frequency spectra exhibit one single component, though broad. Using a sliding window of $16$ points chosen to match the auto-correlation time of passing turbulent structures (typically $1 - 2$ $\mu s$), the signal can be approximated as monochromatic within each window. A single Doppler frequency and its intensity are thus retrieved. Aggregating instantaneous frequency and amplitude estimates across the successive windows then reconstructs the Doppler frequency time evolution with a better time resolution than sliding FFT. Its probability density function (PDF) reproduces the Doppler peak identified in the PSD of the signal very well (see ref.\cite{hennequin2006fluctuation} or \autoref{figapp: music bad snr}). Also, retrieving the amplitude evolution of the dominant frequency allows us to reduce the noise content of the signal at the cost of slightly degrading the time resolution.  \\

As an example, we apply the MUSIC algorithm to discharge $\#82615$ $t=[1-1.1]$s which yields a low signal to noise ratio. The result is displayed in \autoref{fig: comparison music amp}. 

In this example, the raw amplitude correlation function exhibits an irregular shape with sharp transitions and flat regions. The maximum of correlation is $\mathcal{C}<0.7$ and asymmetric on both sides of $\Delta=0$. The low correlation for very small spatial separation is directly linked to the poor signal to noise content of the signal. Applying the MUSIC algorithm restores a well-behaved spatial correlation function. The maximum of the MUSIC-inferred amplitude correlation is close to $\mathcal{C}=0.9$ and is symmetric on both sides of $\Delta=0$. The correlation decreases exponentially with increasing $\Delta$, allowing the correlation length to be estimated (see \autoref{subsection: estimating the correlation lengths}). Note that in this signal, only a single correlation length can be extracted (the flatter decay at $\Delta < - 1$ cm is not representative since too close to the noise level, at about $0.1$).

In the following, the MUSIC algorithm is applied to all discharges, including good signal-to-noise ratio cases, to ensure a consistent treatment across the dataset. For completeness, the analysis was also carried out on the amplitude signal, and the main results of the paper are unchanged. 

\subsection{Estimating the correlation lengths}
\label{subsection: estimating the correlation lengths}

\begin{figure}[h!]
\centering
\begin{minipage}{0.48\textwidth}
    \centering
    \includegraphics[width=\textwidth]{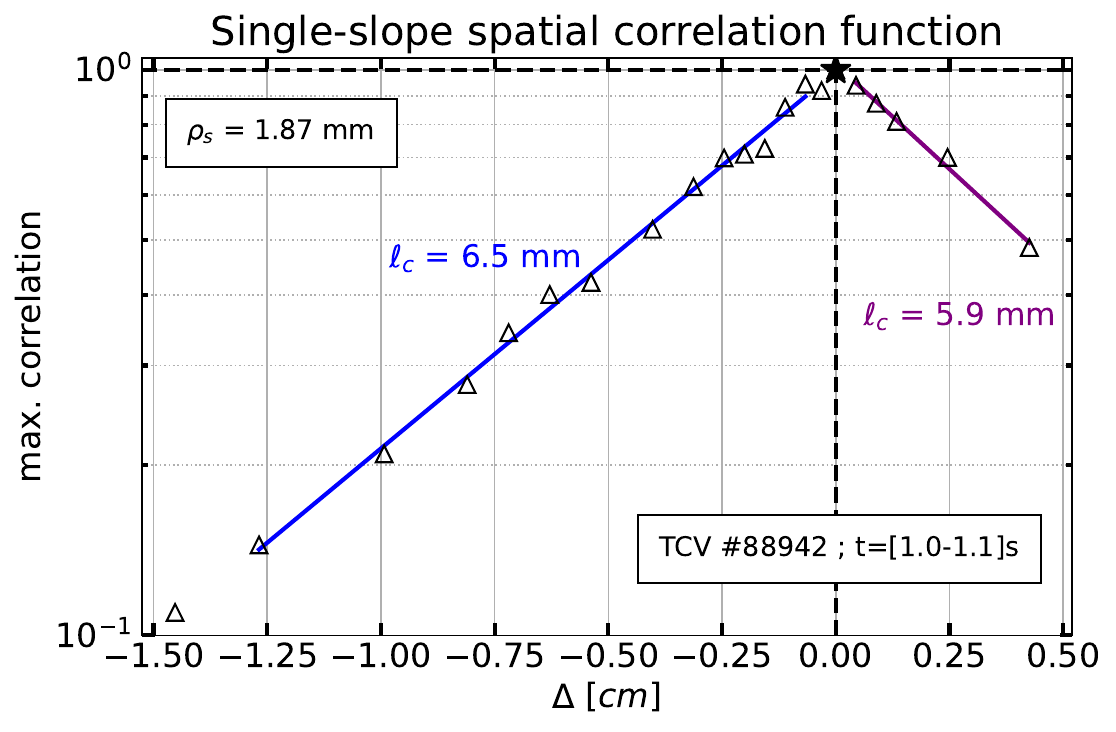}
    \vspace{2mm}
    {\small (a)}
\end{minipage}
\hfill
\begin{minipage}{0.48\textwidth}
    \centering
    \includegraphics[width=\textwidth]{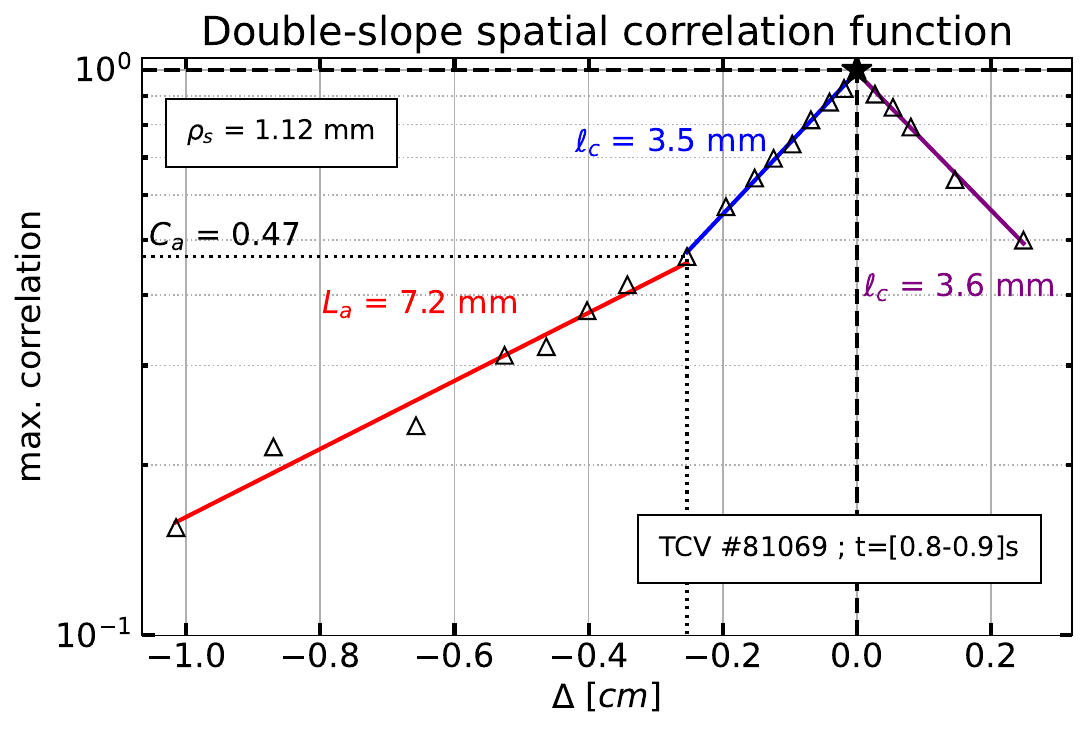}
    \vspace{2mm}
    {\small (b)}
\end{minipage}
\caption[Short Caption]{(a) Spatial correlation function $\#88942$, $t=[1,1.1]$s, $F_{ref}=69$ GHz. (b) Spatial correlation function  $\#81069$, $t=[0.8-0.9]$s, $F_{ref}=60$ GHz.}
\label{fig: example double slope} 
\end{figure}

The slopes of the spatial correlation function are fitted using a least-squares algorithm. The first $\ell_c$ and second $L_a$ lengths are evaluated from the inverse of the slopes. This amounts to considering that the correlation decreases as: $\max (\mathcal{C}) \propto \exp (-\Delta/(\ell_c,L_a))$, on short and large scales, respectively. An example of the fit is shown in \autoref{fig: example double slope} for a case with a single length correlation function and the test case with a two-length correlation function.

In \autoref{fig: example double slope}, at short scale, for positive and negative $\Delta$, the blue and purple fits indicate the first slope of correlation. The correlation function is symmetric, and the two lengths are similar. Note that case (a) indicates a correlation length about two times the one exhibited in case (b). This is a result of the measurement location ($\rho=0.87$ for (a) and $\rho=0.97$ for (b)). When normalising the lengths with the local hybrid Larmor radius $\rho_s = \sqrt{m_i T_e}/(eB)$, this effect is nearly cancelled. At larger $\Delta$, case (b) displays a second slope, in red, about two times the size of the short scale correlation. The short-scale length is commonly used as an estimate of the size of the turbulent structure \cite{mckee2001non, schirmer2007radial}. The second  correlation length at larger distance is likely caused by more radially extended events. We argue, by analogy with simulations (\autoref{section: avalanches simulations}), that it can be interpreted as the signature of avalanche-like transport.  

Finally, the correlation value at which the two slopes separate, $\mathcal{C}_a$, reflects the statistical weight of the long-range component relative to the short-range one. A larger $\mathcal{C}_a$ indicates that the events giving rise to the second slope are become statistically more significant. Two-length spatial correlation function of turbulence have also been identified on ASDEX-Upgrade (AUG) using a similar method. They are presented in ref.\cite{schneider2021overview}.





\subsection{Methodology to account for DBS limitations}

The objective of the current study is to compare short- and long-range correlation lengths under different plasma heating schemes, while keeping otherwise identical plasma conditions. In this first study, the aim is not to provide a quantitative measurement of the correlation lengths -- which would require a DBS angle scan for probing different $k$ combined with synthetic diagnostics -- but rather to investigate their dependencies on specific plasma parameters. 

DBS probes a selected range of perpendicular wavenumbers $k_\perp$, centered around a wavenumber determined by the angle of the incident beam with respect to the plasma cutoff surface. The measurements therefore only probe a specific region of the turbulence spectrum. To compare the different discharges, the primary requirement is that $k_\perp$ remains consistent across all measurements. As the probing angle is kept fixed for discharges with similar equilibrium, both the reference and hopping channels sample similar perpendicular wavenumbers. Across all discharges, $k_\perp \in [5.4 - 8]$ cm$^{-1}$ with variations arising mainly from differences in measurement depth. The few exceptions are excluded from the analysis.

The measurements may fall in either the linear or nonlinear scattering regime depending on the fluctuation amplitude \cite{gusakov2002non, gusakov2004spatial, krutkin2019nonlinear}. In the nonlinear case, multiple scattering contributions are competing: the backscattered signal (BS) at the cutoff layer which can be analysed from the linear theory; and multiple forward scattering (FS) contributions along the beam path, particularly in the edge region at high fluctuation amplitude. If the FS contributions are too important, the resulting signal is poorly localized. This leads to a wider Doppler frequency spectrum and makes the estimation of both the Doppler shift and the correlation difficult. To assess in which regime the measurements fall in, we rely on the standard nonlinear criterion established analytically in \cite{gusakov2002non, gusakov2005multiple} assuming linear squared refractive index and Gaussian wavenumber spectrum,
\begin{align}
    \gamma = G^2 \frac{\omega_o^2}{c^2}l_{c}x_c \left(\frac{\delta n}{n_c}\right)^2 \ln \frac{x_c}{l_{c}} 
    \label{eq: nonlinear criterion}
\end{align}
Where $\omega_0$ is the probing pulsation, $c$ the speed of light, $x_c$ the distance to the cutoff layer,  $l_{c}$ the radial correlation length, and $\delta n $ the density fluctuation level  normalized to the cutoff density $n_c$. 
$G$ is a correcting factor taking into account the beam polarisation. $\gamma \lesssim 1$ corresponds to the linear regime and $\gamma \gg 1$ to the nonlinear regime. In ref.\cite{gusakov2004analytical}, it is considered that at $\gamma \sim 20$, the FS contributions become larger than the BS one. 

The evaluation of $\gamma$ for the range of probing frequencies $\lesssim 60$ $GHz$, typical correlation length $l_{c}=$  3-\SI{5}{mm} and $\delta n/n = 1-5\%$ gives $\gamma \sim 3-10$. This places the measurements in the weakly nonlinear regime, where nonlinear effects are expected to remain limited. A more accurate evaluation of the criterion can be obtained by integrating the local value of the criterion along the beampath provided by the raytracing code, taking into account the local optical index from the measured density and magnetic field profiles. This estimate yields the same range or slightly lower $\gamma$. We expect then that the collected data fall in the linear or weakly nonlinear regime and are therefore assumed to be mostly localized at the cutoff location. 
However, to assess more precisely the nonlinear response of the DBS diagnostic, and enable quantitative measurements, full-wave modelling will be required, involving fluctuation level profile measurements and/or evaluation from gyrokinetic simulations.

\section{Comparison with short pulse reflectometry}
\label{section: comparison with}

The DBS measures at a specific $k_\perp \sim 6-10$ cm$^{-1}$. To estimate the influence of the perpendicular wavenumber on the estimation of the correlation length, we compare with the short pulse reflectometer (SPR) that measures at a much lower $k_\perp$. SPR estimates density fluctuations by measuring the flight time of pulses sent towards the cutoff layer at a normal incidence \cite{krutkin2023method}. The SPR system \cite{molina2019vband} shares the launcher system with the DBS and provides measurement from the same upper lateral port, although at normal angle to the plasma surface. 

\begin{figure}[h]
	\centering
	\includegraphics[width=0.5\textwidth]{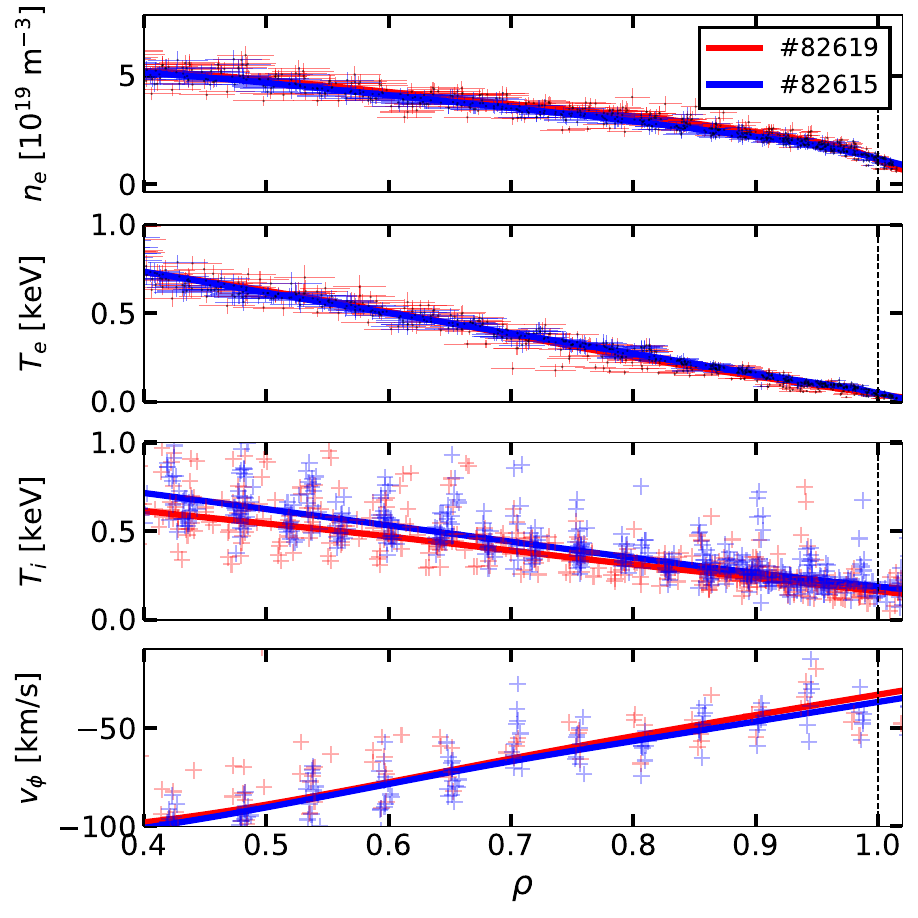}
	\caption{Comparison of kinetic profiles as a function of the normalised poloidal flux coordinate $\rho$ for DBS shot $\#82615$ $t=[1.4,1.6]$s and SPR shot $\#82619$ $t=[1.5,1.7]$s. From top to bottom: electron density $n_e$, electron temperature $T_e$, ion temperature $T_i$ and toroidal velocity $v_\phi$.}
	\label{fig: dbs spr profile comparison}
\end{figure}

To compare SPR and DBS we repeat shot $\#82615$ at NBH=\SI{300}{kW}. The matched profiles are shown in \autoref{fig: dbs spr profile comparison}. The electron density $n_e$ and electron temperature $T_e$ are computed using TCV's Thomson scattering system \cite{blanchard2019thomson}. The ion temperature $T_i$ and toroidal velocity $v_\phi$ estimated for carbon VI impurities are measured with the CXRS system \cite{thesisBagnoto, marini2017thesis}.  

The kinetic profiles are well matched for the time window in which the comparison is performed. However, we point out that most of discharge $\#82619$ is subject to MHD events, possibly due to magnetic islands. Therefore, the comparison has only been made at the end of the discharge, after the gas fueling has stopped. \\ 

\begin{figure}[h]
	\centering
	\includegraphics[width=0.5\textwidth]{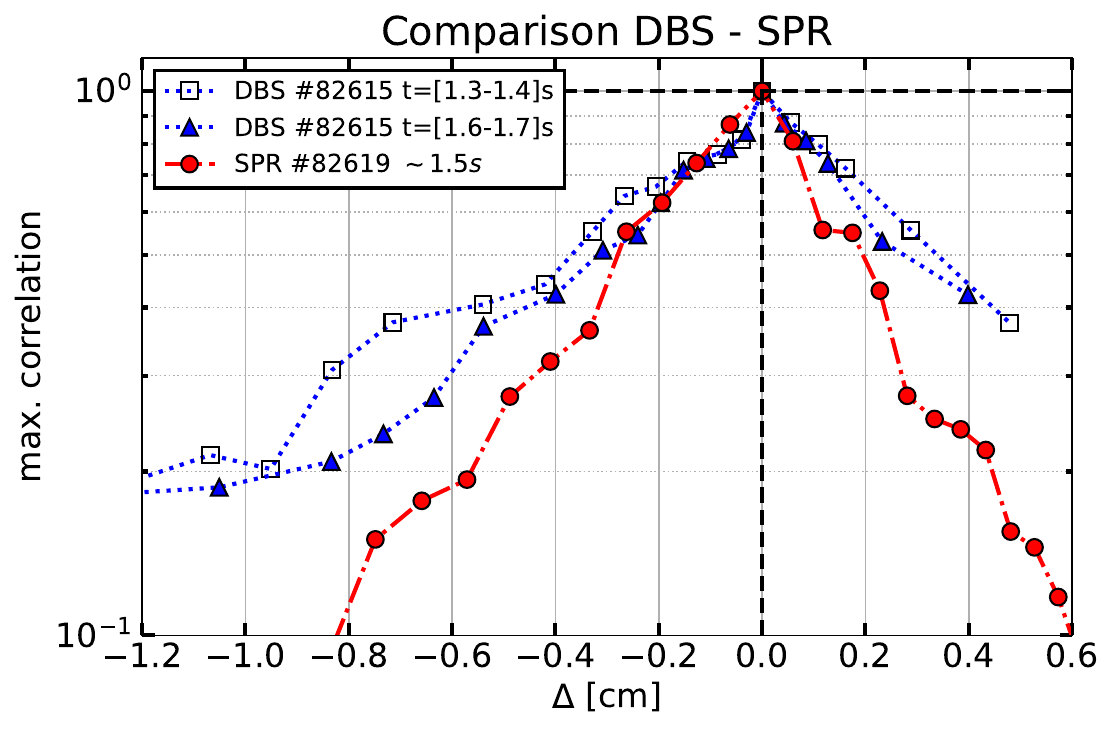}
	\caption{Comparison of spatial correlation functions obtained with DBS and with SPR. Both diagnostics measure the correlation length for $\rho$ between 0.93 and 0.96.}
	\label{fig: comparison correlation spr dbs}
\end{figure}

The correlation analysis with the SPR is performed following the method described in \cite{krutkin2024validation}. The comparison of the spatial correlation function computed with DBS and SPR is performed for the first time and shown in \autoref{fig: comparison correlation spr dbs}. 

At short scale, both correlation functions display a similar decay, although the DBS correlation function is symmetric whereas the SPR correlation function is not. The comparable decay indicates a similar estimation of the turbulence correlation length. At larger scale, the DBS correlation functions still display the same exponential decay, and no second slope of correlation is found on this example. Comparatively, the SPR correlation decay faster after $\Delta \sim$ \SI{0.3}{cm}. This can be an effect of the nonlinear response of SPR diagnostic as already detailed in \cite{krutkin2023method}. 

We underline that this comparison is intended as a first step, but that fullwave computation should be performed for both diagnostics to understand the effects leading to the different correlation functions at larger scales. This will be part of a future work, together with additional diagnostic comparison.  



\section{Scaling of correlation lengths with heating and perpendicular velocity}
\label{section: scaling of}

The aim is to change the plasma kinetic profiles so as to modify the distance to the instability threshold and / or change the dominant instability. Therefore, we use the same equilibrium configuration with different heating schemes and power. The analysis is done for L-mode Deuterium plasmas. To be sure not to transit into H-mode even at larger NBH power, the chosen configuration is an unfavourable $B\times \nabla B$ corresponding to a higher L-H power threshold. The magnetic equilibrium is shown in \autoref{fig: ex beamtracing correlation} together with the beamtracing for shot $\#81069$. This shape has proven to be very stable in terms of MHD activity and plasma breakdown. 
We use a co-current neutral beam injection (NBI-1) among the two available neutral beams. \\ 

Each correlation measurement takes approximately $\SI{100}{ms}$, as shown in \autoref{fig: frequency pattern}. To maximise the available measurement time within each discharge, two heating steps are used, each lasting $400$ or $600$ $ms$, allowing several correlation measurements at different radial locations to be performed per step. A persistent challenge in these experiments -- notably for NBH plasmas -- is density control, which is particularly critical for DBS since the plasma density profile determines the measurement location. The operational window for combined NBH and ECH at TCV is quite narrow, partly because of the density control. For these reasons, ECH-only and NBH-only have been analysed separately. 

Several approaches to improve density control were tested over the course of the campaign. For shot numbers below $\#82607$, an ohmic phase was introduced between the two heating steps, allowing the density to decrease towards its set value. For shot numbers above $\#88942$, real-time density control was implemented using the RAPDENS code \cite{pastore2026applicationsnovelmodelbasedrealtime}. \\ 

\begin{figure}[h]
	\centering
	\includegraphics[width=0.8\textwidth]{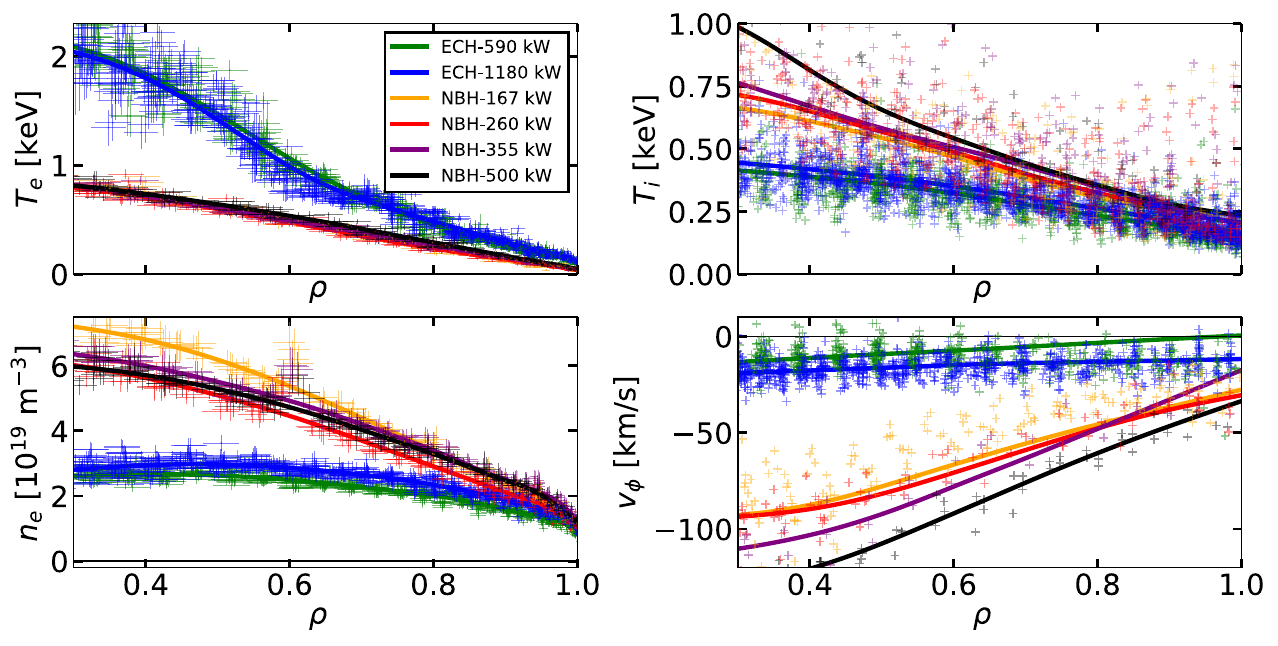}
	\caption{Kinetic profiles used for the linear analysis as a function of the normalised poloidal flux coordinate $\rho$. The plasma profiles are taken at $t=[1.65-1.75] ; [0.85 - 0.96] ; [0.65 - 0.75]$ for $NBH = 167, 260, 355, 500$ $kW$ respectively. Plasma profiles are taken at $t=[0.8-1.1] ; [1.3-1.7]$ for $ECH=590, 1180$ $kW$.}
	\label{fig: profile gene}
\end{figure}

The injected NBH power is scanned from $140$ $kW$ to $685$ kW while the ECH power ranges from $590$ to $1180$ kW. The summary of the performed discharges, alongside the time of each heating plateau is presented in \autoref{table: performed deuterium tcv shot} of \autoref{appendix: summary of the performed discharges}. Some shots have been repeated to perform correlation measurements using the SPR diagnostic. The typical resulting kinetic profiles are presented in \autoref{fig: profile gene} for a certain number of heating powers. Six heating plateaus have been chosen at $P_{NBH} = 167, 260, 355, 500$ $kW$ and $P_{ECH} = 590, 1180$ $kW$. On the first left column are presented the electron temperature $T_e$ and electron density $n_e$. On the right column are displayed the ion temperature $T_i$ and toroidal velocity $v_\phi$ estimated for carbon VI impurities.

Compared with the NBH case, the ECH scenario leads to a larger $T_e$ and a comparatively lower $T_i$. The feedback-controlled electron density is lower in ECH plasmas so as to better couple to the cyclotron wave. The NBH scenario leads to $T_i \approx T_e$ and to an important toroidal velocity. Note that $v_\phi$ in the core is negative because NBI-1 is co-current ($I_p<0$). Although increasing NBH power leads to a modification of the ion temperature profile, no important modification is observed for $T_e$. Furthermore, the profiles are modified in the core, but the edge gradients ($\rho \approx 0.8 - 1$) barely increase with heating power. The main differences come from the density profile and the toroidal velocity, which increases with the NBH power. 

Based on the above profiles, a linear estimation of the growth rate and frequency has been obtained using GENE by taking gradient values at $\rho = 0.95$. In each tested case, the dominant instability is found to rotate in the electron diamagnetic direction. It is assumed that trapped electron modes (TEM) is the main instability as it is often the case for TCV plasmas \cite{bottino2006linear, camenen2007impact}. One of the initial objectives of the study was to modify the profiles sufficiently to change the dominant instability. While this may not have been achieved, a wide range of heating profiles was explored, allowing us to probe the stiffness of the plasma \cite{garbet2004profile}. This is of particular interest, as stiffness is directly related to transport and may be related to the long-range correlation analysed in this contribution. \\

\begin{figure}[h]
	\centering
	\includegraphics[width=0.5\textwidth]{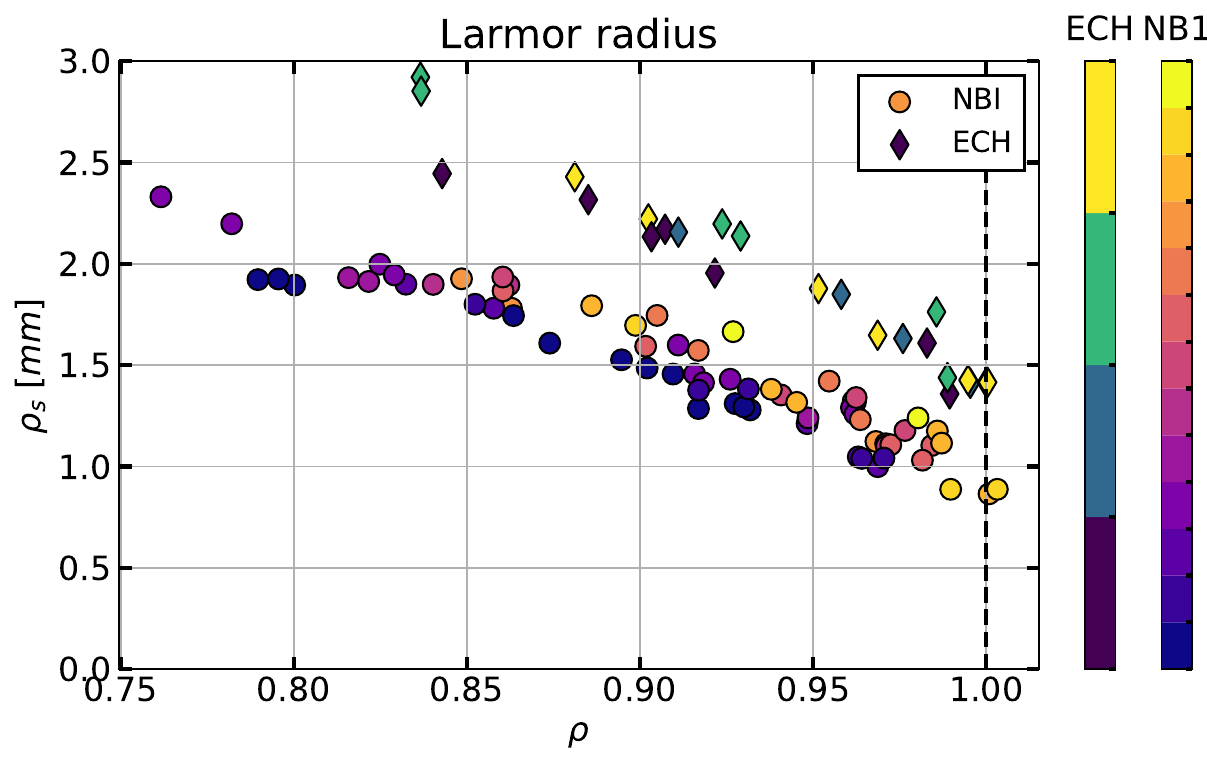}
	\caption{Larmor radius $\rho_s = \sqrt{m_i T_e}/(eB)$ as a function of the normalised radius $\rho$ and heating for discharges in \autoref{table: performed deuterium tcv shot}. ECH ranges from $590$ to $1180$ $kW$ and NB1 from $140$ to $685$ $kW$. }
	\label{fig: rho rhos}
\end{figure}

To compare the turbulence correlation lengths with characteristic scales, we evaluate the hybrid Larmor radius $\rho_s$ at the position of the reference DBS channel for each heating plateau. The results are shown in \autoref{fig: rho rhos} as a function of the normalised radius $\rho$. The color indicates the NBH and ECH heating power.  

Consistently with the profiles presented in \autoref{fig: profile gene}, $\rho_s$ varies of about $10-20$ $\%$ as a function of the heating power. In ECH cases, $T_e$ and consequently $\rho_s$ tend to be larger. The small variation of $\rho_s$ despite an important change in heating power is the result of the profiles being very stiff in the experiments. The local value of the Larmor radius is used herafter to normalise the correlation lengths $\ell_c$ and $L_a$.

\subsection{Dependence on NBH power}

The correlation lengths are analysed for all ECH and NBH shots listed in \autoref{table: performed deuterium tcv shot}. While most ECH shots provide correlation data from the core region ($\rho \sim 0.86$) up to the separatrix, the NBH cases generally only provide data up to $\rho \sim 0.98$, with only two correlation functions extending beyond this radial position. Therefore, in this section, we first analyse the dependence of the correlation lengths on NBH power. The role of velocity shear in the reduction of turbulent structures is then examined using the ECH shots in \autoref{subsection: reduced structures in the Er well}. Differences between the ECH and NBH cases are discussed where relevant. \\

\begin{figure}[h!]
\centering
\begin{minipage}{0.48\textwidth}
    \centering
    \includegraphics[width=\textwidth]{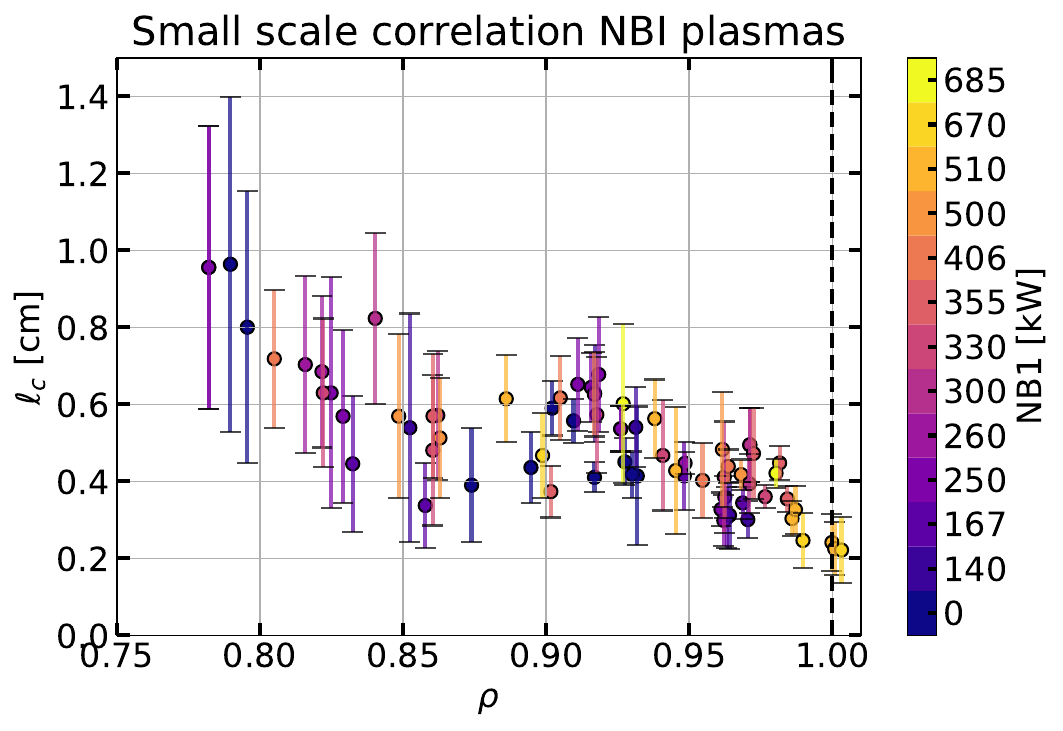}
    \vspace{2mm}
    {\small (a)}
\end{minipage}
\hfill
\begin{minipage}{0.48\textwidth}
    \centering
    \includegraphics[width=\textwidth]{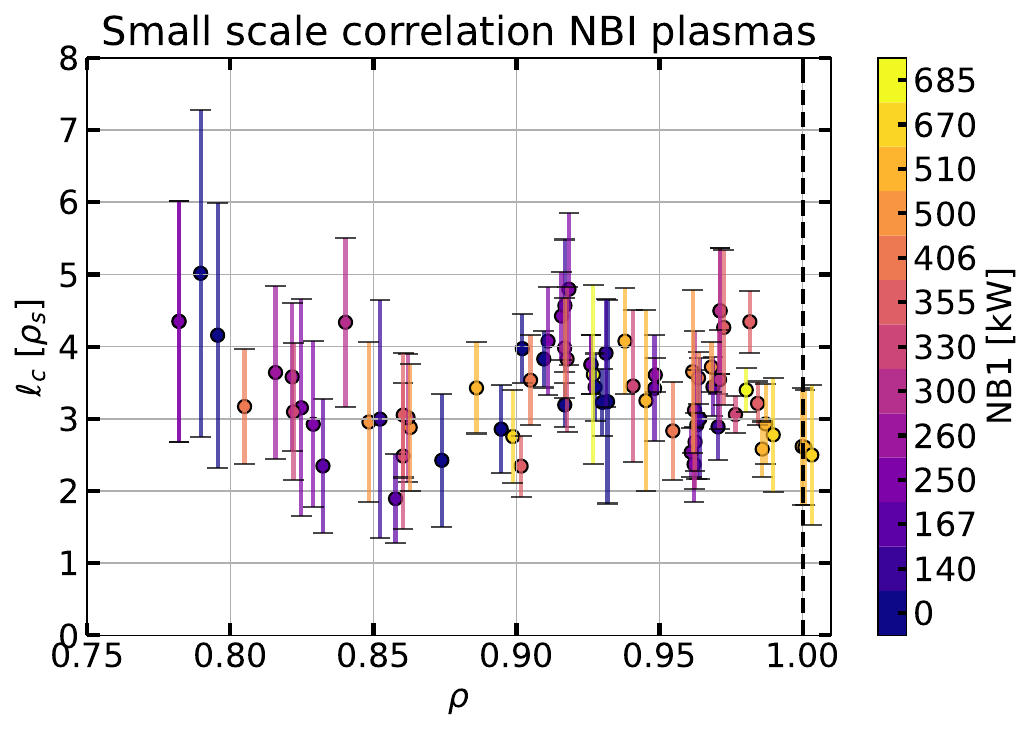}
    \vspace{2mm}
    {\small (b)}
\end{minipage}
\caption[Short Caption]{Spatial correlation length of the turbulent structures at short distance $\ell_c$ as a function of $\rho$ and heating power. (a) In centimeters. (b) Normalised using the local hybrid Larmor radius $\rho_s$. The errorbar on $\ell_c$ is computed using the difference between the two estimates (each side of $\mathcal{C} = 1$) plus the standard deviation of the fit.}
\label{fig: lc nbi cm rhos}
\end{figure}

The short scale correlation $\ell_c$ is shown in \autoref{fig: lc nbi cm rhos} for NBH  cases as a function of $\rho$ and of the heating power. On the left are shown the lengths in $cm$, on the right they are normalised using the local hybrid Larmor radius $\rho_s$. 

The size of the turbulent structures is found to decrease towards the edge of the plasma. When renormalising with $\rho_s$ this effect is effectively nearly cancelled. The linear dependence of $\ell_c$ on $\rho_s$ and/or $\rho_i$ was already reported in earlier contributions \cite{kurzan2000measurement, mckee2001non}. The short scale correlation lengths lie between $2$ and $6$ $\rho_s$ with no clear dependence on the heating power. For ECH cases -- shown in \autoref{fig: lc la vperp ech} -- the short scale correlation length is roughly constant in the core at $\ell_c \sim 3-4$ $\rho_s$. A small reduction of $\ell_c$ is found closer to the separatrix. This may be an effect of the radial electric field $E_r$ well. It is further analysed together with the ECH cases in \autoref{subsection: reduced structures in the Er well}. The short scale correlation lengths are in agreement with previous studies on other tokamaks. On ASDEX-Upgrade (AUG), using DBS, the turbulence correlation length is found slightly larger at $\ell_c^{AUG} \approx 6-8$ $\rho_i$ in ref.\cite{schneider2021overview} and slightly lower at $\ell_c^{AUG}\approx 3$ $\rho_i$ in ref.\cite{kurzan2000measurement}. Another contribution reports on the correlation length on the order of $0.3$ to $2$ $cm$ in the $\rho \in [0.7-1]$ range but does not normalise with the local Larmor radius \cite{schirmer2007radial}. On DIII-D, ref.\cite{mckee2001non} using beam-emission spectroscopy (BES) reports a radial correlation of $\ell_c^{DIII-D} \approx 5$ $\rho_i$. 

The size of the turbulent structures is not expected to scale with the heating power. Instead, it should reflect the type of turbulence that dominates the measurement region. In our experiments it appears that the turbulence is always dominated by TEM, hence we do not expect a change in the size of $\ell_c$. Ref.\cite{schneider2021overview} states that AUG L-mode turbulence is dominated by ITG/TEM in the core and becomes dominated by collisional drift waves closer to the edge. They measure the correlation length at $\rho=0.7$, therefore a similar correlation length is consistent with this view. Additionally, the turbulence correlation length depends on the underlying magnetic \cite{fedorczak2013dynamics} and velocity \cite{Biglari1990} shears. The magnetic shear is set by the geometry and is constant in the presented experiments. The role of the velocity shear is detailed in \autoref{subsection: reduced structures in the Er well}.
\\

\begin{figure}[h]
	\centering
	\includegraphics[width=0.5\textwidth]{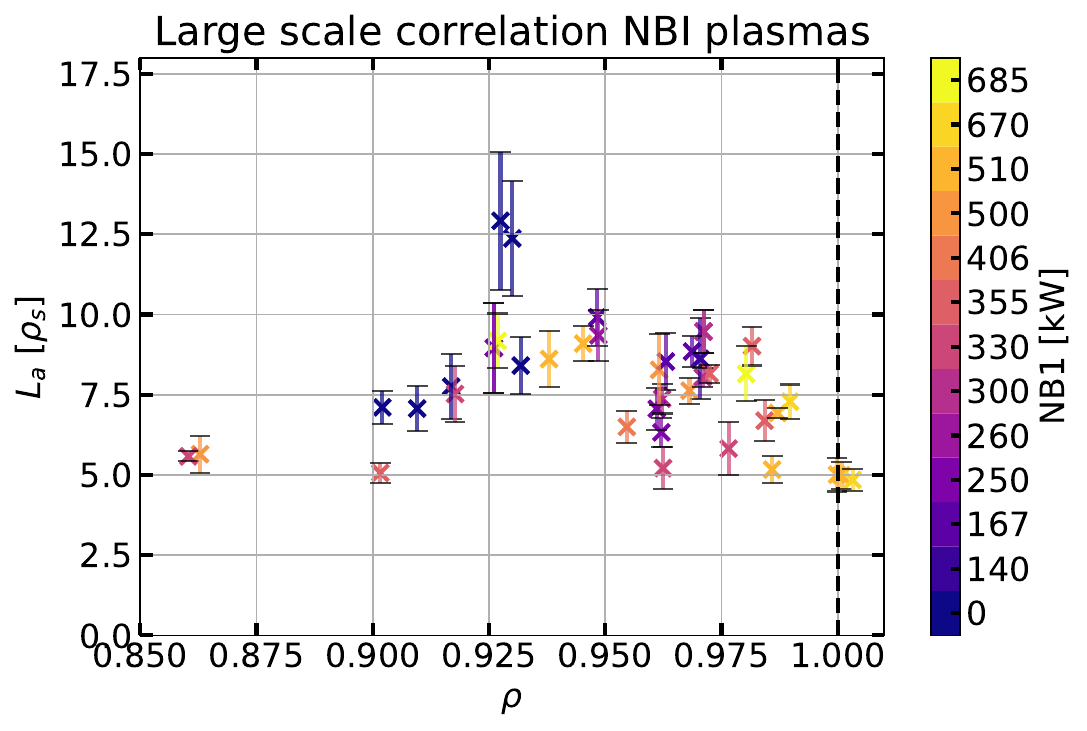}
	\caption{Spatial correlation length of the turbulence at large scale $L_a$ normalised by $\rho_s$ as a function of $\rho$ and heating power.}
	\label{fig: la nbi rhos}
\end{figure}

The second correlation length, at larger distance, is shown in \autoref{fig: la nbi rhos}. The statistics is reduced because some spatial correlation functions only display a single slope, interpreted as a short scale correlation. 

Large scale correlations range from $5$ to $15$ $\rho_s$ and are found mostly for $\rho > 0.9$ for all heating powers. At equal $\rho$, cases at lower power and in particular the Ohmic heated shots exhibit a larger second correlation length. 
For non-ohmic cases, no clear dependence on NBH power is found on the large scale correlation length. The measured lengths are found to increase from $\rho=0.9$ to $\rho=0.95$ and decrease afterwards towards the separatrix. Note that two points seem to be located on / after $\rho=1$. While the reference may be very close to the separatrix, the correlation lengths are mostly measured inwards (\autoref{fig: example double slope}). Therefore, the lengths should be understood as being inside the last closed flux surface. The large scale correlation identified in \cite{schneider2021overview} is $L_a \approx 2.7$ cm ($\approx 15-20$ $\rho_s$), which is larger than the lengths reported here. However, their measurements were performed deeper into the plasma, at $\rho = 0.7$. \\

\begin{figure}[h]
	\centering
	\includegraphics[width=0.5\textwidth]{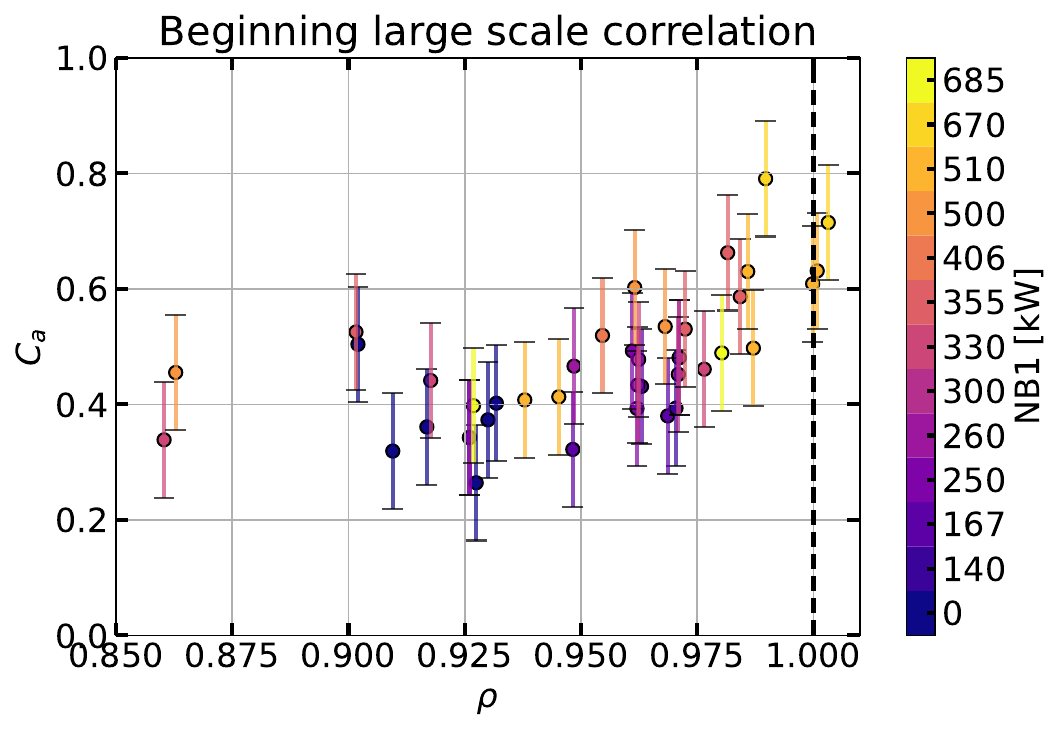}
	\caption{Importance of the large scale correlation length with respect to the small scale correlation. The error is estimated at $0.1$ for all cases after careful consideration of the spatial correlation functions.}
	\label{fig: ca nbi rhos}
\end{figure}

Finally, an important indicator of the weight of the second correlation slope relative to the first is where the break between the two slopes is located. This is estimated with $C_a$ the "beginning" of the second correlation slope (see \autoref{fig: example double slope}). In \autoref{fig: ca nbi rhos}, $C_a$ is displayed as a function of $\rho$ and the NBH power. 

Two main comments can be made on \autoref{fig: ca nbi rhos}. First, $C_a$ increases from $\rho \sim 0.925$ to the separatrix. Second, a slight increase of $C_a$ is observed with respect to the heating power for the same $\rho$. This indicates that the turbulence events leading to a second correlation slope are more important at larger power and close to the edge of the confined plasma. 


\subsection{Reduced structures in the $E_r$ well}
\label{subsection: reduced structures in the Er well}

\begin{figure}[h]
	\centering
	\includegraphics[width=0.6\textwidth]{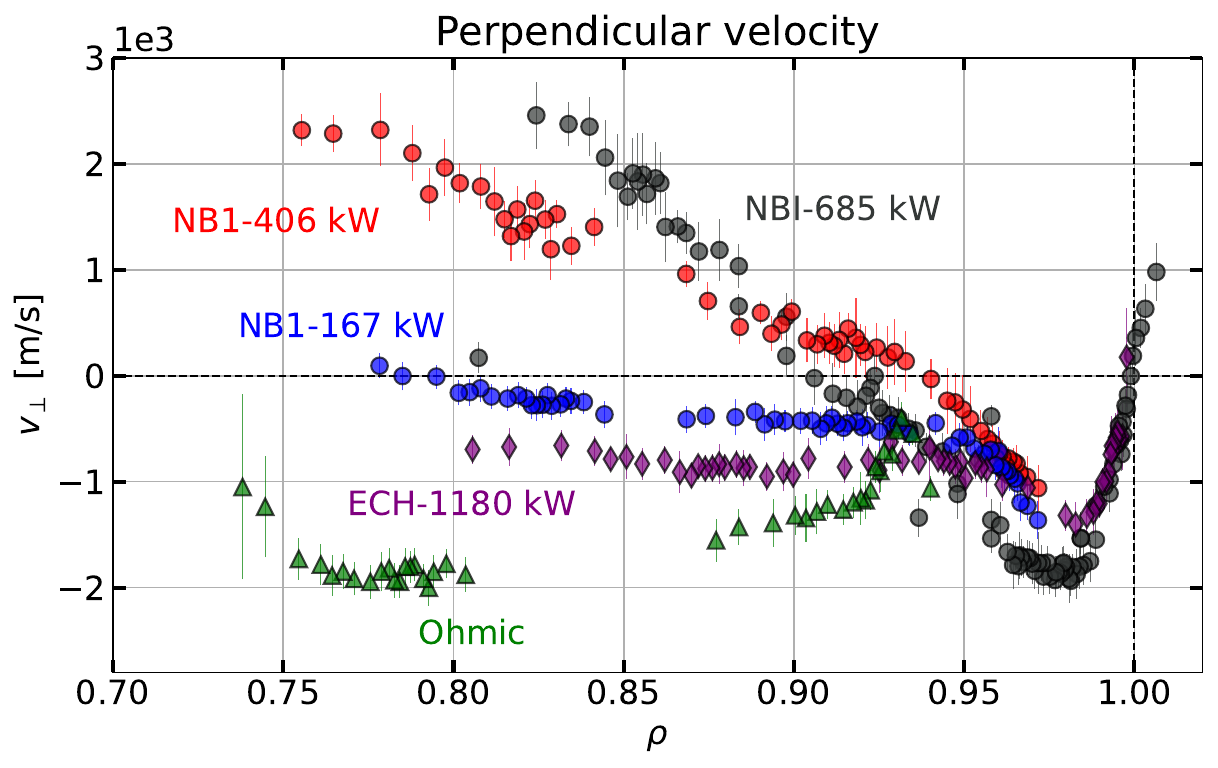}
	\caption{Perpendicular velocity profiles taken from the hopping channel as a function of $\rho$ for different heating schemes (NBI and ECH) and powers. From top to bottom the cases are taken from shot $88945$ , $82615$, $82607$, $82611$ and $81065$.}
	\label{fig: vperp ohmic ech nbi}
\end{figure}

This section examines the link between the perpendicular velocity shear and the size of the turbulent structures. According to the BDT model \cite{Biglari1990}, a sufficiently large shear stretches and decorrelates turbulent eddies, thereby reducing their characteristic size. DBS provides access to the perpendicular velocity $v_\perp$. A large variety of profiles is obtained depending on the type of heating. A selection of $v_\perp$ profiles is shown in \autoref{fig: vperp ohmic ech nbi}. 

The perpendicular velocity profile is found to be articulated in two parts, possibly the sign of different physics: the $E_r$ well region and the core region. Close to the separatrix, in the $E_r$ well region, a similar profile is found regardless of the chosen heating. Note that the NBH case at \SI{685}{kW} leads to a slightly wider $E_r$ well compared to the others. A previous study on TCV \cite{rienacker2025survey} already reported that the $E_r$ well varies little in diverted L-mode discharges for a given magnetic geometry. Interestingly, in some cases, a "bump" is observed between the $E_r$ well region and the core region. This is also described in ref.\cite{rienacker2025survey}. Here, it is particularly visible on the Ohmic case but it also occurs in some ECH or NBH discharges without a clear trend in heating scheme or power. Further inside the core, the $v_\perp \approx \langle E_r \rangle / B$ behaviour is in agreement with the radial force balance, 
\begin{align}
   \langle E_r \rangle = \frac{\nabla p_i}{n e} + v_\phi B_\theta - v_\theta B_\phi \; , 
\end{align}
where $p_i$ is the ion pressure, $n$ is the density, $e$ the electron charge, $v_\phi$ the toroidal velocity and $v_\theta$ the poloidal velocity. The role of the toroidal rotation is particularly noticeable. In the present scenario, The poloidal and toroidal magnetic fields $B_\theta$ and $B_\phi$ are negative, NBH-1 is co-current and $\mathbf{I}_p<0$, so that $v_\phi B_\theta$ is positive. Simultaneously, NBH also increases the ion pressure, which translates into a more negative $\nabla p_i$. Both effects of NBH act in opposite directions regarding the sign of $v_\perp$. In the TCV shots reported here, it appears that the toroidal rotation effect is stronger at large NBH power, ultimately leading to a positive $v_\perp$ in the core. \\

\begin{figure}[h]
	\centering
	\includegraphics[width=0.5\textwidth]{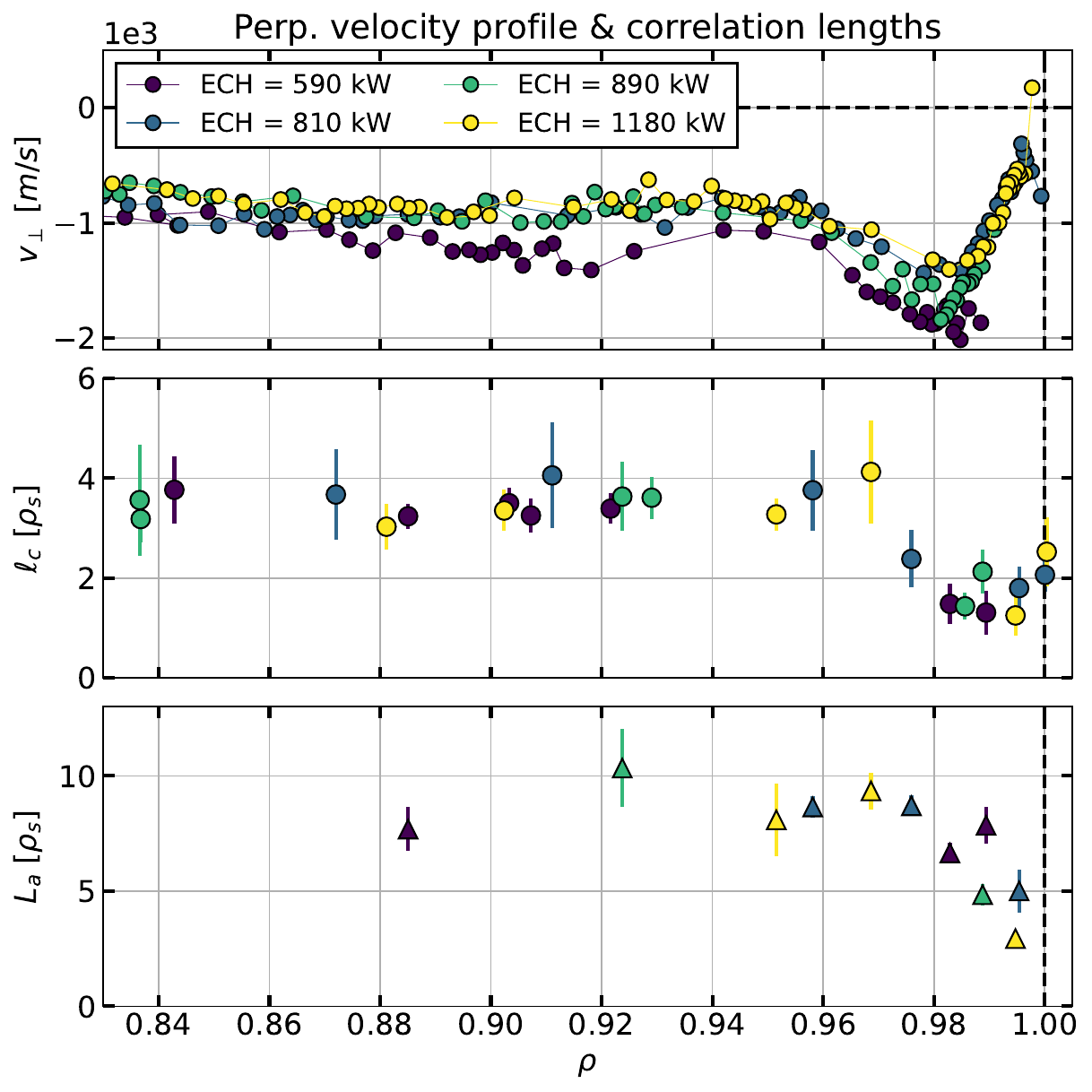}
	\caption{Comparison between perpendicular velocity profiles and correlation lengths. (top) $v_\perp$ profiles for $4$ ECH powers. (middle) Short range correlation length $\ell_c$. (bottom) Long range correlation length $L_a$.}
	\label{fig: lc la vperp ech}
\end{figure}

The ECH cases lead to comparable perpendicular velocity profiles for the four heating power values. The $v_\perp$ profiles are plotted against the short and long range correlation lengths in \autoref{fig: lc la vperp ech}. 

In \autoref{fig: lc la vperp ech}, a shallow $E_r$ is found for all cases. In the core, the perpendicular velocity is constant at $v_\perp \sim - 1$ $km.s^{-1}$. It is found that in the region of large shear corresponding to the $E_r$ well, both $\ell_c$ and $L_a$ are significantly reduced. The short range correlation length $\ell_c$ is reduced roughly by a factor $2$. \\ 

The reduction of the correlation lengths close to the separatrix can be a consequence of the strong velocity shear in that region. The reduction is more pronounced for ECH cases than for the few NBH cases that have data in the $E_r$ well. A notable difference between the two cases is how wide the well is, which directly impact the velocity shear that is visible by the turbulent structures. Indeed, for NBH cases, the $E_r$ well is larger than for ECH cases. In particular, it is wider than the structures themselves, so they may be able to live quite unperturbed in this region. For ECH, the $E_r$ well is quite narrow and the structures are efficiently sheared by the perpendicular velocity. This may also indicate that the perpendicular velocity shear in the core is not sufficiently large to reduce the size of the turbulent structures, even at high NBH power. 

An alternative interpretation of the reduction in structure size inside the $E_r$ well is that it may result from a diagnostic effect. Indeed, the DBS nonlinearity criterion (\autoref{eq: nonlinear criterion}) depends on the turbulence intensity $\delta n / n$ which increases close to the separatrix. More importantly, the turbulence correlation function may be narrower when measured in the nonlinear regime of turbulence \cite{krutkin2020investigation}. A dedicated study of this effect with the help of fullwave simulations is planned for future work.

\section{Two-length correlation functions in simulations}
\label{section: avalanches simulations}

In order to get better insight on the experimental correlation lengths, we compute the correlation on two simulations performed with the reduced nonlinear flux-driven code Tokam1D \cite{panico2025importance}. The goal here is to compare simulation with and without avalanches and to apply an analysis procedure analogous to that used for the experiments. In particular, we aim to determine whether the presence of two-length correlation functions is related to avalanche transport.  

The Tokam1D model considers an electrostatic isothermal plasma with hot ions and assumes no scale separation between the flux surface averaged and fluctuating quantities. It features two instabilities: collisional drift waves (CDW) and interchange, governed through an effective conductivity $C$ and gravity $g$ parameters. The TEM instability is not included in the model; however, it belongs to the broader family of interchange instabilities. The model is reduced to one-dimension by projecting the fluctuations onto a single parallel and poloidal wave vectors. This reduction makes the model fast enough for simulations on confinement timescales while still resolving small turbulence scales.

\begin{figure}[h!]
\centering
\begin{minipage}{0.48\textwidth}
    \centering
    \includegraphics[width=\textwidth]{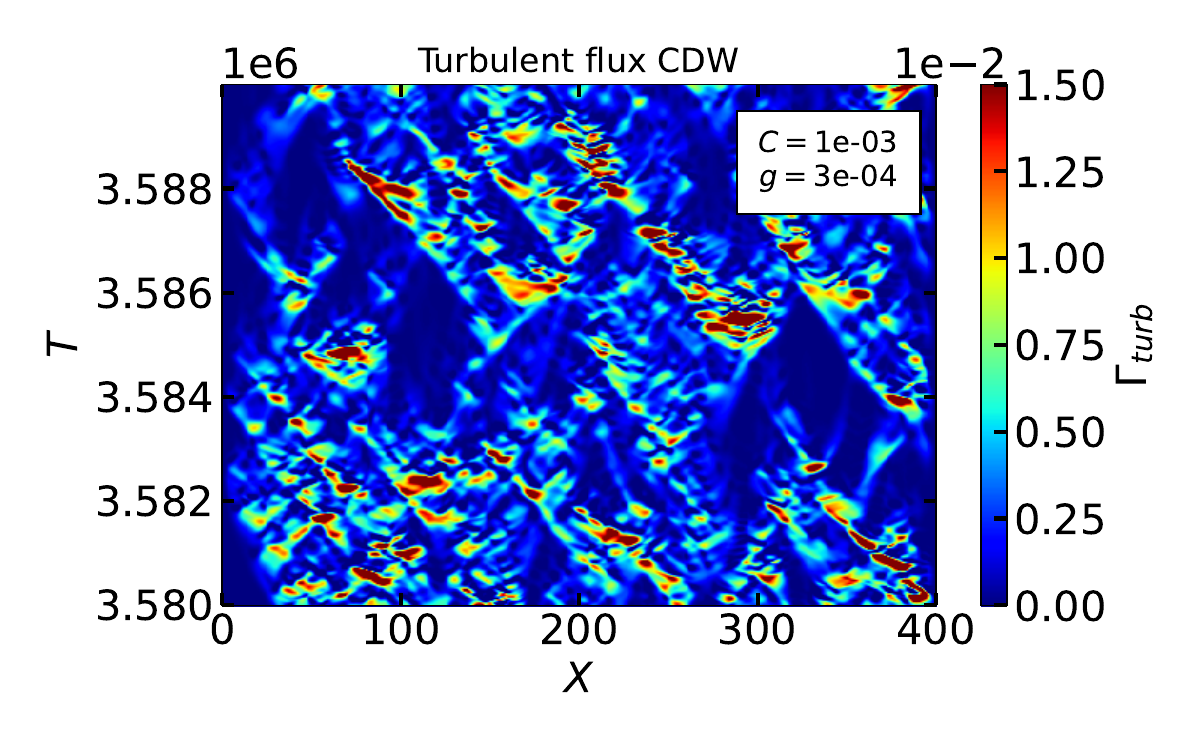}
    \vspace{2mm}
    {\small (a)}
\end{minipage}
\hfill
\begin{minipage}{0.48\textwidth}
    \centering
    \includegraphics[width=\textwidth]{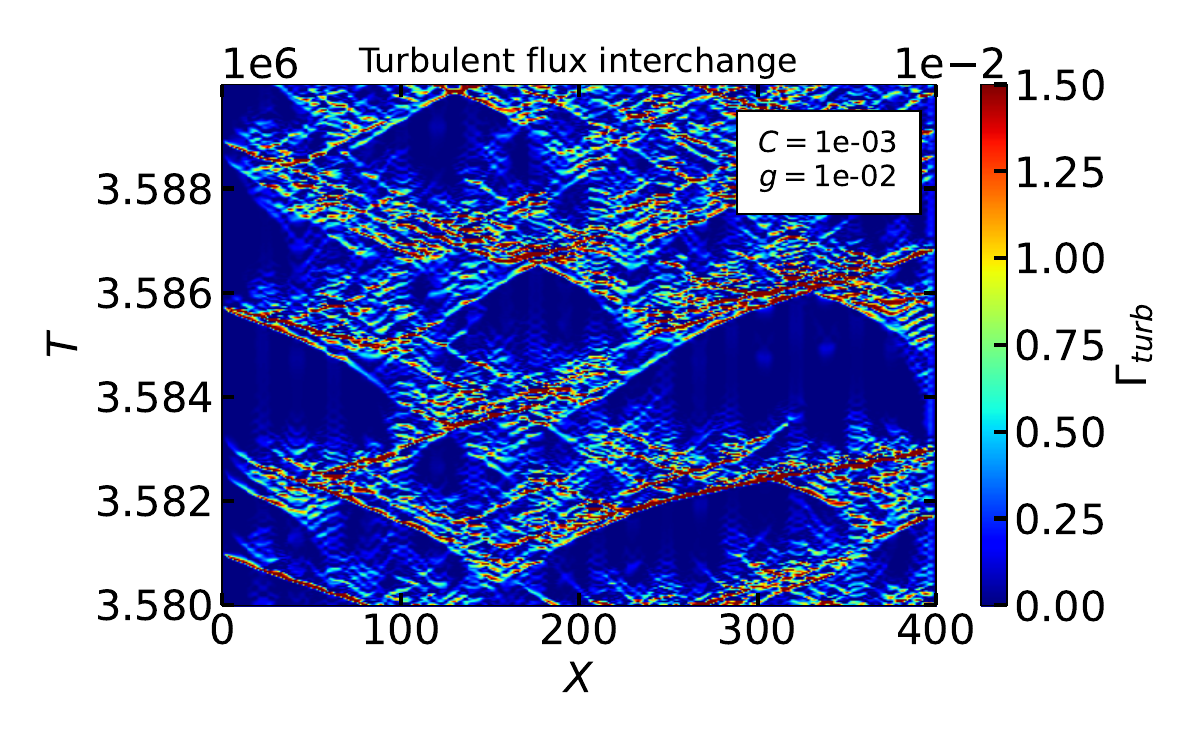}
    \vspace{2mm}
    {\small (b)}
\end{minipage}
\caption[Short Caption]{Examples of turbulent particle flux 
$\Gamma_{turb} = -2k_y \Im(N_k \phi_k^*)$ at steady state.
(a) $(C,g)=(10^{-3}, 3 \times 10^{-4})$. 
(b) $(C,g)=(10^{-3}, 10^{-2})$. Both computed with a fixed source at $S_N(0)=10^{-4}$.}
\label{fig:flux_turb_transition_avalanches}
\end{figure}

Here we compare two simulations, already introduced in ref.\cite{panico2025generation}, at low and large interchange drive. Their particle fluxes are presented in \autoref{fig:flux_turb_transition_avalanches} as a function of the time $T$ normalized to the ion-cyclotron frequency $\omega_{ci} = eB/m_i$ and radial direction $X$ normalized to the hybrid Larmor radius $\rho_s = m_i c_s / (e B)$, where $e$ is the elementary charge, $B$ the magnetic field, $m_i$ the ion mass and $c_s=\sqrt{T_e /m_i}$ the sound speed.

\begin{figure}[h!]
\centering
\begin{minipage}{0.48\textwidth}
    \centering
    \includegraphics[width=\textwidth]{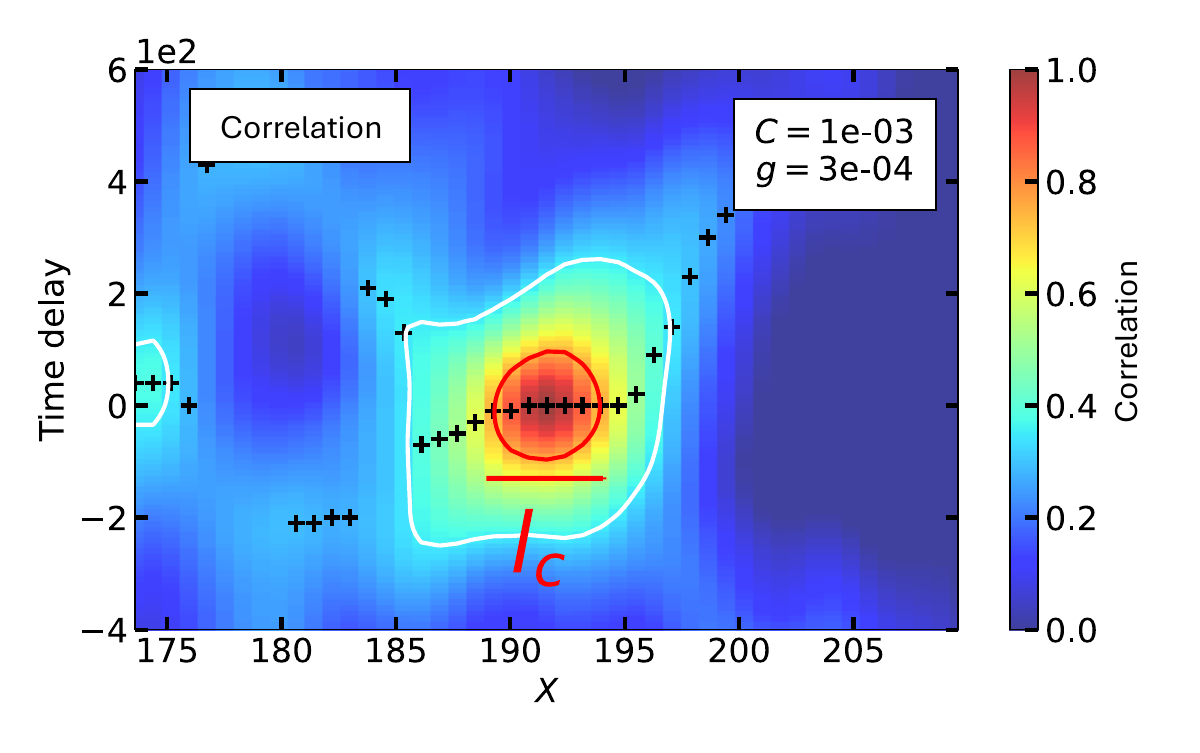}
    \vspace{2mm}
    {\small (a)}
\end{minipage}
\hfill
\begin{minipage}{0.48\textwidth}
    \centering
    \includegraphics[width=\textwidth]{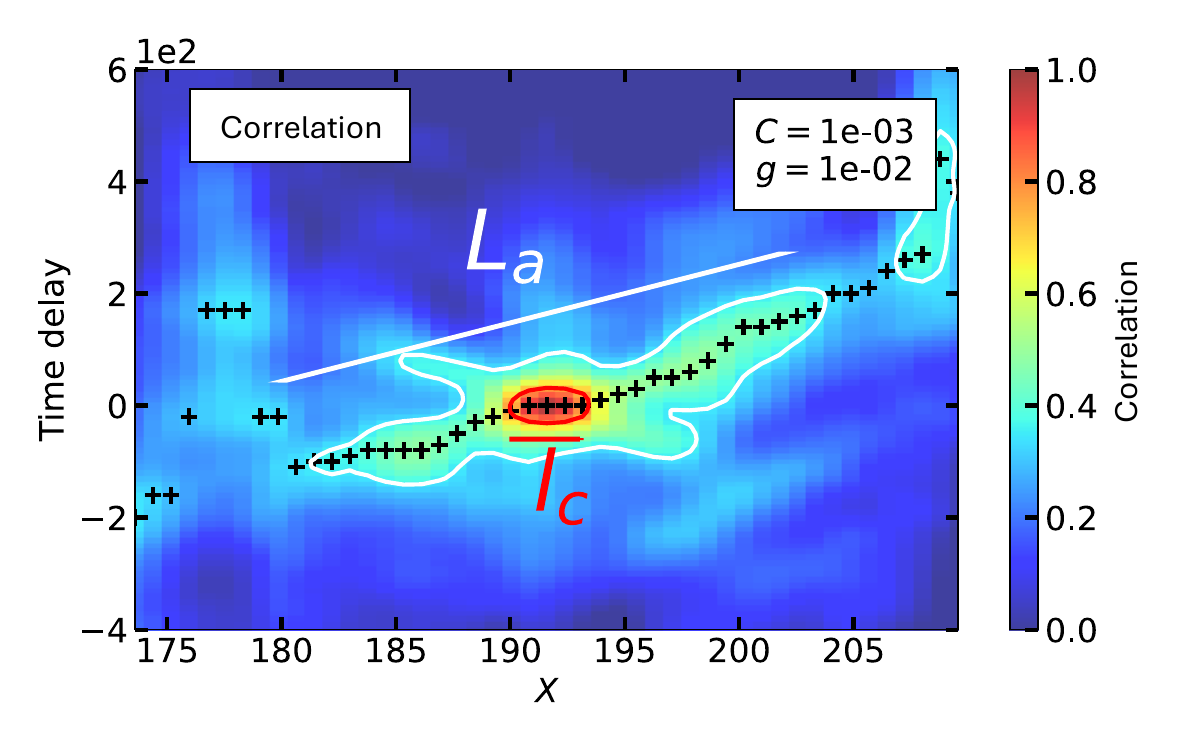}
    \vspace{2mm}
    {\small (b)}
\end{minipage}
\caption[Short Caption]{Radial correlation function of density fluctuations amplitude versus time delay and radial direction. (a) $(C,g)=(10^{-3}, 3 \times 10^{-4})$. 
(b) $(C,g)=(10^{-3}, 10^{-2})$.}
\label{fig: simulation 2d correlation function}
\end{figure}

In \autoref{fig:flux_turb_transition_avalanches} (a), the  turbulent flux of particles is local both in space and time in the form of small bursts. In \autoref{fig:flux_turb_transition_avalanches} (b), the turbulent flux exhibits diagonal stripes across large portions of the simulation domain. Those correspond to ballistic transport events of particles resembling avalanches. To compare the simulations with DBS data, we compute the spatial correlation function of density fluctuations amplitude on both cases. An example is shown in \autoref{fig: simulation 2d correlation function} for the two cases described above. The correlation is computed between a reference $X=192$ slice and the neighboring slices $X\pm \Delta X$. The color represents the correlation, the black crosses indicate the time delays at which the correlation is maximal. Two contour lines delimit $\mathcal{C} = 0.75$ (red) and $\mathcal{C} = 1/e$ (white). 

\begin{figure}[h!]
	\centering
	\includegraphics[width=0.5\textwidth]{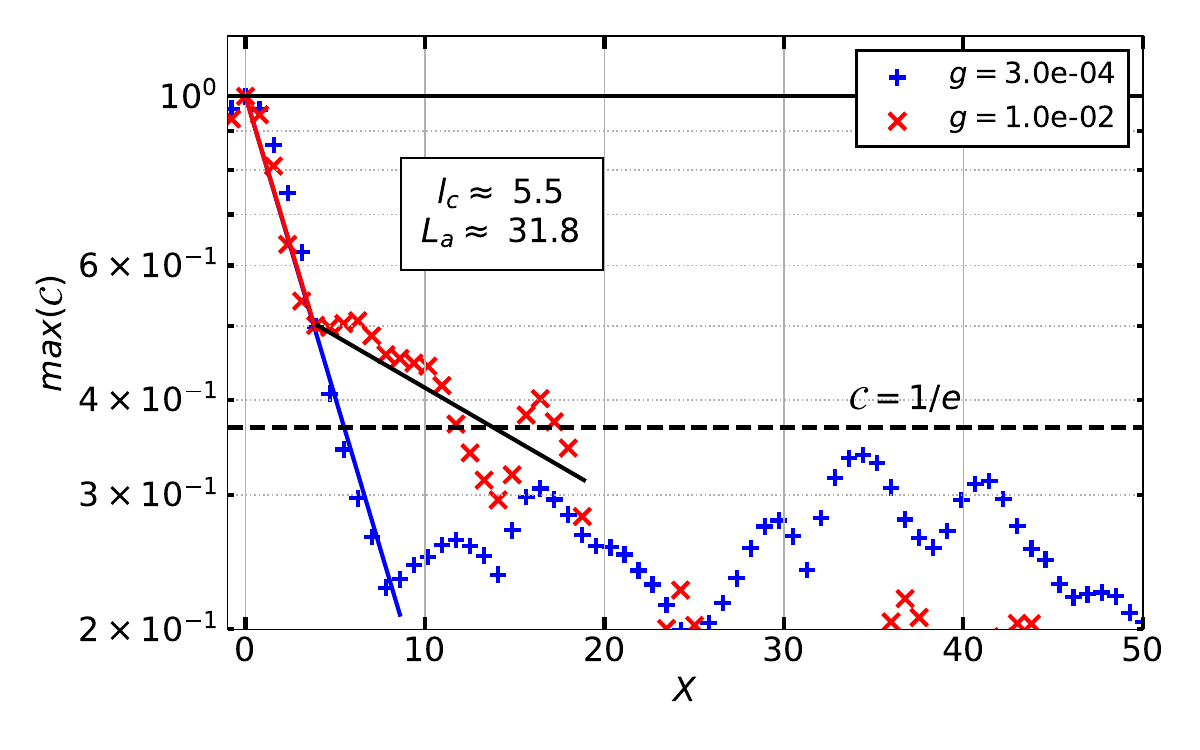}
	\caption{Radial correlation function computed from the maximum of coherence. (a) No avalanches: g = $3 \times 10^{-4}$. (b) Avalanche dominated: $g = 10^{-2}$. Both taken at $C = 10^{-3}$.}
	\label{fig: simulation 1d spatial correlation function}
\end{figure}

Consistently with \autoref{fig:flux_turb_transition_avalanches} (a), the correlation in \autoref{fig: simulation 2d correlation function} (a) displays a single, non-propagating, structure.  Its typical size of a few $\rho_s$ is labelled $l_c$. In the case dominated by avalanche transport, two characteristic lengths appear. The smaller, similarly to the first case, displays no time delay. The larger is tilted diagonally, indicating a propagating event. The size and tilt of $L_a$ are comparable to the observed avalanches in the turbulent flux \autoref{fig:flux_turb_transition_avalanches} (b). The avalanche dominated case thus displays two slopes for the correlation function. This is best seen when plotting the maximum of coherence at each position, see \autoref{fig: simulation 1d spatial correlation function}.

In \autoref{fig: simulation 1d spatial correlation function}, the double slope of correlation appears clearly in the case at large $g$, dominated by avalanches, only. The fits are performed on the logarithm of the radial correlation, assuming an e-folding decay $e^{-x/l_c}$. The short-scale correlation lengths are similar for the two cases at $l_c \approx 5.5 ~\rho_s$. The large-scale correlation length is roughly equal to $L_a \approx 32 ~\rho_s$, which is consistent with \autoref{fig:flux_turb_transition_avalanches}.  \\

This kind of two-length correlation functions when the transport is dominated by avalanches was already described in ref.\cite{dif2017b} (figure 3) for gyrokinetic simulations. In those simulations, the structures are found at $\ell_c = 6-8$ $\rho_s$ and the avalanches at $L_a \approx 40$ $\rho_s$. While this is consistent with the reduced model estimation of $L_a \approx 30$ $\rho_s$, it is much larger than the experimental estimation of $L_a \approx 5-15$ $\rho_s$. The size discrepancy between experiments and TCV measurements is not yet fully understood. Further experiments under different turbulence regimes, together with dedicated simulations of TCV, are needed to investigate the origin of this discrepancy.

\section{Conclusion}
\label{section: discussion}

Spatial correlation measurements of density fluctuations have been performed in the TCV tokamak using Doppler backscattering (DBS). $14$ shots have been selected at different heating powers in both ECH and NBH. This corresponds to $115$ spatial correlation functions ranging from $\rho=0.75$ to $\rho=0.99$. The signal was preprocessed using the MUSIC algorithm, resulting in reduced noise content and a better estimation of correlation lengths. \\

In about half of the cases, two distinct spatial scales were identified in the correlation functions: a short-range decay associated with the size of the turbulent structures and a longer-range component manifesting as a second correlation slope. The two-length correlation functions also appear in the simulations, as demonstrated in the reduced model Tokam1D. More importantly, they appear when the transport is bursty and avalanche-like. In simulations, the second length provides a good estimate of the extension of avalanche events. By analogy with the simulations, one can interpret the two-length correlation function measured with DBS as an indication of avalanche transport in TCV. \\

At short scales, $\ell_c$ ranges from $3$ to $5$ hybrid Larmor radii $\rho_s$. In the $E_r$ well region, the short scale correlation goes down to $1.5$ $\rho_s$. At large scale, the second correlation length ranges from $5$ to $15$ $\rho_s$. The method described in the current contribution has also been used in ref.\cite{schneider2021overview} for AUG data where it is found that the correlation lengths at the pedestal top $\rho=0.7-0.9$ are of the range of $\ell_c \sim 5$ $\rho_s$ and $L_a \sim 15-20$ $\rho_s$. The measurements furthest into the plasma in the present work, at $\rho=0.77$, are in the same ballpark at $\ell_c = 4.5 - 5$ $\rho_s$. However, in our case, no second correlation length is found at these locations. \\

In simulations, the structure size is of the order of $5$ $\rho_s$ while the avalanches extension is much larger of the order of $30-40$ $\rho_s$. The difference in avalanche lengths between experiments and simulations may be related to the relatively small size of TCV, which has a minor radius of \SI{25}{cm}. The limited plasma size may prevent the formation or accommodation of large-scale avalanches within the available plasma volume. \\

The break between the two correlation-length scaling regimes, $C_a$, is used as an indicator of the relative contribution of long-range events compared with short-range events. It is found that $C_a$ increases towards the plasma edge and with increasing heating power. This effect is particularly pronounced in NBH plasmas. \\

Both the short- and large- scale correlation lengths are found to be reduced in the ECH $E_r$ well region. Interestingly, the reduction is observed only in ECH plasmas and not in the few NBH cases for which data is available in this region. A possible reason is that the NBH $E_r$ well is wider than its ECH counterpart. Therefore, the turbulent structures experience a smaller average shear in the region and are not as stretched and decorrelated as for the ECH plasmas. The large scale correlation length is also found to decrease in the $E_r$ well region. An alternative interpretation of this reduction is given by a nonlinear behaviour of the DBS in this particular region. Investigating the role of the nonlinear criterion will be part of future work. \\

DBS has previously been used to characterise the correlation length of turbulent structures \cite{schirmer2007radial}. Its application to the identification of long-range correlation events was first introduced in ref.\cite{schneider2021overview} for AUG data. This work presents the first measurements on TCV and comparison with simulations. However, because the DBS diagnostic selects a specific perpendicular wavenumber $k_\perp$, the correlation lengths reported here should not be interpreted as absolute measures of the structure size. They are most meaningful when compared with measurements performed at similar $k_\perp$, although a similar length is also obtained with SPR at low $k_\perp$. A more complete characterization of the structure size would require a scan of the DBS angle, complemented by a full-wave synthetic diagnostic. In addition, further experimental investigations covering other turbulence regimes and larger devices are needed to compare with simulations. \\

Finally, two future directions are envisioned. First, the same measurements have also been performed in hydrogen and deuterium plasmas with matched profiles, providing an opportunity to investigate isotope effects on the correlation lengths. Second, the correlation analysis could be extended to discharges immediately preceding the L-H transition. This would make it possible to assess whether the measured correlation lengths are consistent with the turbulence suppression criterion.

\newpage
%

\ack{Sample text inserted for demonstration.}

\funding{Sample text inserted for demonstration.}

\roles{Sample text inserted for demonstration.}

\data{Sample text inserted for demonstration.}

\suppdata{Sample text inserted for demonstration.}

\bibliographystyle{unsrt}
\bibliography{references.bib}

@article{bak1987self,
	title={Self-organized criticality: An explanation of the 1/f noise},
	author={Bak, Per and Tang, Chao and Wiesenfeld, Kurt},
	journal={Physical review letters},
	volume={59},
	number={4},
	pages={381},
	year={1987},
	publisher={APS}
}

@article{Biglari1990,
  title={Influence of sheared poloidal rotation on edge turbulence},
  author={Biglari, Hamed and Diamond, PH and Terry, PW},
  journal={Physics of Fluids B: Plasma Physics},
  volume={2},
  number={1},
  pages={1--4},
  year={1990},
  publisher={American Institute of Physics}
}

@article{diamond1995dynamics,
	title={On the dynamics of turbulent transport near marginal stability},
	author={Diamond, Patrick H and Hahm, TS},
	journal={Physics of Plasmas},
	volume={2},
	number={10},
	pages={3640--3649},
	year={1995},
	publisher={American Institute of Physics}
}

@article{carreras1996model,
	title={A model realization of self-organized criticality for plasma confinement},
	author={Carreras, BA and Newman, D and Lynch, VE and Diamond, PH},
	journal={Physics of Plasmas},
	volume={3},
	number={8},
	pages={2903--2911},
	year={1996},
	publisher={American Institute of Physics}
}

@article{mazzucato1998microwave,
	title={Microwave reflectometry for magnetically confined plasmas},
	author={Mazzucato, E},
	journal={Review of Scientific Instruments},
	volume={69},
	number={6},
	pages={2201--2217},
	year={1998},
	publisher={American Institute of Physics}
}

@article{sarazin1998intermittent,
  title={Intermittent particle transport in two-dimensional edge turbulence},
  author={Sarazin, Y and Ghendrih, Ph},
  journal={Physics of Plasmas},
  volume={5},
  number={12},
  pages={4214--4228},
  year={1998},
  publisher={American Institute of Physics}
}

@article{kurzan2000measurement,
	title={Measurement and scaling of the radial correlation lengths of turbulence at the plasma edge of ASDEX Upgrade},
	author={Kurzan, B and de Pe{\~n}a Hempel, S and Holzhauer, E and Scott, B and Serra, F and Suttrop, W and Zeiler, A and ASDEX Upgrade Team and others},
	journal={Plasma physics and controlled fusion},
	volume={42},
	number={3},
	pages={237},
	year={2000},
	publisher={IOP Publishing}
}

@article{politzer2000observation,
	title={Observation of avalanchelike phenomena in a magnetically confined plasma},
	author={Politzer, PA},
	journal={Physical review letters},
	volume={84},
	number={6},
	pages={1192},
	year={2000},
	publisher={APS}
}

@article{sarazin2000transport,
	title={Transport due to front propagation in tokamaks},
	author={Sarazin, Y and Garbet, X and Ghendrih, Ph and Benkadda, S},
	journal={Physics of Plasmas},
	volume={7},
	number={4},
	pages={1085--1088},
	year={2000},
	publisher={American Institute of Physics}
}

@article{estrada2001turbulence,
	title={Turbulence and beam size effects on reflectometry measurements},
	author={Estrada, T and Sanchez, J and Zhuravlev, V and de la Luna, E and Branas, B},
	journal={Physics of Plasmas},
	volume={8},
	number={6},
	pages={2657--2665},
	year={2001},
	publisher={American Institute of Physics}
}

@article{hirsch2001doppler,
	title={Doppler reflectometry for the investigation of propagating density perturbations},
	author={Hirsch, M and Holzhauer, E and Baldzuhn, J and Kurzan, B and Scott, B},
	journal={Plasma physics and controlled fusion},
	volume={43},
	number={12},
	pages={1641},
	year={2001},
	publisher={IOP Publishing}
}

@article{mckee2001non,
  title={Non-dimensional scaling of turbulence characteristics and turbulent diffusivity},
  author={McKee, GR and Petty, CC and Waltz, RE and Fenzi, C and Fonck, RJ and Kinsey, JE and Luce, TC and Burrell, KH and Baker, DR and Doyle, EJ and others},
  journal={Nuclear Fusion},
  volume={41},
  number={9},
  pages={1235},
  year={2001},
  publisher={IOP Publishing}
}

@article{conway2004plasma,
	title={Plasma rotation profile measurements using Doppler reflectometry},
	author={Conway, GD and Schirmer, J and Klenge, S and Suttrop, W and Holzhauer, E and ASDEX Upgrade Team and others},
	journal={Plasma Physics and Controlled Fusion},
	volume={46},
	number={6},
	pages={951},
	year={2004},
	publisher={IOP Publishing}
}

@article{hennequin2004doppler,
	title={Doppler backscattering system for measuring fluctuations and their perpendicular velocity on Tore Supra},
	author={Hennequin, P and Honor{\'e}, C and Truc, A and Qu{\'e}m{\'e}neur, A and Lemoine, N and Chareau, J-M and Sabot, R},
	journal={Review of Scientific Instruments},
	volume={75},
	number={10},
	pages={3881--3883},
	year={2004},
	publisher={AIP Publishing}
}

@article{bottino2006linear,
	title={Linear stability analysis of microinstabilities in electron internal transport barrier non-inductive discharges},
	author={Bottino, A and Sauter, O and Camenen, Y and Fable, E},
	journal={Plasma physics and controlled fusion},
	volume={48},
	number={2},
	pages={215},
	year={2006},
	publisher={IOP Publishing}
}

@article{hennequin2006fluctuation,
	title={Fluctuation spectra and velocity profile from Doppler backscattering on Tore Supra},
	author={Hennequin, P and Honor{\'e}, C and Truc, A and Qu{\'e}m{\'e}neur, A and Fenzi-Bonizec, C and Bourdelle, C and Garbet, X and Hoang, GT and Tore Supra team and others},
	journal={Nuclear fusion},
	volume={46},
	number={9},
	pages={S771},
	year={2006},
	publisher={IOP Publishing}
}

@article{honore2006quasi,
	title={Quasi-optical Gaussian beam tracing to evaluate Doppler backscattering conditions},
	author={Honor{\'e}, C and Hennequin, P and Truc, A and Qu{\'e}m{\'e}neur, A},
	journal={Nuclear fusion},
	volume={46},
	number={9},
	pages={S809},
	year={2006},
	publisher={IOP Publishing}
}

@article{schirmer2006radial,
	title={The radial electric field and its associated shear in the ASDEX Upgrade tokamak},
	author={Schirmer, J and Conway, GD and Zohm, H and Suttrop, W and ASDEX Upgrade Team and others},
	journal={Nuclear fusion},
	volume={46},
	number={9},
	pages={S780},
	year={2006},
	publisher={IOP Publishing}
}

@article{camenen2007impact,
	title={Impact of plasma triangularity and collisionality on electron heat transport in TCV L-mode plasmas},
	author={Camenen, Y and Pochelon, A and Behn, R and Bottino, A and Bortolon, A and Coda, S and Karpushov, A and Sauter, O and Zhuang, G and others},
	journal={Nuclear fusion},
	volume={47},
	number={7},
	pages={510},
	year={2007},
	publisher={IOP Publishing}
}

@article{schirmer2007radial,
  title={Radial correlation length measurements on ASDEX Upgrade using correlation Doppler reflectometry},
  author={Schirmer, J and Conway, GD and Holzhauer, E and Suttrop, W and Zohm, H and ASDEX Upgrade Team},
  journal={Plasma Physics and Controlled Fusion},
  volume={49},
  number={7},
  pages={1019--1039},
  year={2007}
}

@article{idomura2009study,
	title={Study of ion turbulent transport and profile formations using global gyrokinetic full-f Vlasov simulation},
	author={Idomura, Yasuhiro and Urano, H and Aiba, Nobuyuki and Tokuda, S},
	journal={Nuclear Fusion},
	volume={49},
	number={6},
	pages={065029},
	year={2009},
	publisher={IOP Publishing}
}

@article{ku2009full,
	title={Full-f gyrokinetic particle simulation of centrally heated global ITG turbulence from magnetic axis to edge pedestal top in a realistic tokamak geometry},
	author={Ku, Susan and Chang, Choong-Seock and Diamond, Patrick H},
	journal={Nuclear Fusion},
	volume={49},
	number={11},
	pages={115021},
	year={2009},
	publisher={IOP Publishing}
}

@article{mcmillan2009avalanchelike,
	title={Avalanchelike bursts in global gyrokinetic simulations},
	author={McMillan, Ben F and Jolliet, S and Tran, TM and Villard, L and Bottino, A and Angelino, P},
	journal={Physics of Plasmas},
	volume={16},
	number={2},
	year={2009},
	publisher={AIP Publishing}
}

@article{dif2010validity,
	title={On the validity of the local diffusive paradigm in turbulent plasma transport},
	author={Dif-Pradalier, Guilhem and Diamond, PH and Grandgirard, Virginie and Sarazin, Yanick and Abiteboul, J and Garbet, Xavier and Ghendrih, Ph and Strugarek, A and Ku, S and Chang, CS},
	journal={Physical Review E—Statistical, Nonlinear, and Soft Matter Physics},
	volume={82},
	number={2},
	pages={025401},
	year={2010},
	publisher={APS}
}

@article{vermare2012detection,
	title={Detection of geodesic acoustic mode oscillations, using multiple signal classification analysis of Doppler backscattering signal on Tore Supra},
	author={Vermare, Laure and Hennequin, Pascale and G{\"u}rcan, {\"O}zg{\"u}r D and Tore Supra Team and others},
	journal={Nuclear Fusion},
	volume={52},
	number={6},
	pages={063008},
	year={2012},
	publisher={IOP Publishing}
}

@article{fedorczak2013dynamics,
	title={Dynamics of tilted eddies in a transversal flow at the edge of tokamak plasmas and the consequences for L--H transition},
	author={Fedorczak, N and Ghendrih, Ph and Hennequin, Pascale and Tynan, GR and Diamond, PH and Manz, P},
	journal={Plasma Physics and Controlled Fusion},
	volume={55},
	number={12},
	pages={124024},
	year={2013},
	publisher={IOP Publishing}
}

@article{dif2017b,
	title={The E$\times$ B staircase of magnetised plasmas},
	author={Dif-Pradalier, Guilhem and Hornung, G and Garbet, X and Ghendrih, Ph and Grandgirard, V and Latu, G and Sarazin, Y},
	journal={Nuclear Fusion},
	volume={57},
	number={6},
	pages={066026},
	year={2017},
	publisher={IOP Publishing}
}

@article{pinzon2019experimental,
	title={Experimental investigation of the tilt angle of turbulent structures in the core of fusion plasmas},
	author={Pinz{\'o}n, JR and Happel, T and Hennequin, Pascale and Angioni, Clemente and Estrada, Teresa and Lebschy, Alexander and Stroth, Ulrich and ASDEX Upgrade Team and others},
	journal={Nuclear Fusion},
	volume={59},
	number={7},
	pages={074002},
	year={2019},
	publisher={IOP Publishing}
}

@article{pinzon2019measurement,
	title={Measurement of the tilt angle of turbulent structures in magnetically confined plasmas using Doppler reflectometry},
	author={Pinz{\'o}n, JR and Estrada, Teresa and Happel, T and Hennequin, Pascale and Blanco, E and Stroth, Ulrich and Teams, TJ-II and others},
	journal={Plasma Physics and Controlled Fusion},
	volume={61},
	number={10},
	pages={105009},
	year={2019},
	publisher={IOP Publishing}
}

@article{schneider2021overview,
  title={Overview of the isotope effects in the ASDEX Upgrade tokamak},
  author={Schneider, PA and Hennequin, Pascale and Bonanomi, N and Dunne, M and Conway, GD and Plank, U and ASDEX Upgrade Team and EUROfusion MST1 Team and others},
  journal={Plasma Physics and Controlled Fusion},
  volume={63},
  number={6},
  pages={064006},
  year={2021},
  publisher={IOP Publishing}
}

@article{ghendrih2022role,
doi = {10.1088/1742-6596/2397/1/012018},
url = {https://dx.doi.org/10.1088/1742-6596/2397/1/012018},
year = {2022},
month = {dec},
publisher = {IOP Publishing},
volume = {2397},
number = {1},
pages = {012018},
author = {Philippe Ghendrih and Guilhem Dif-Pradalier and Olivier Panico and Yanick Sarazin and Hugo Bufferand and Guido Ciraolo and Peter Donnel and Nicolas Fedorczak and Xavier Garbet and Virginie Grandgirard and Pascale Hennequin and Eric Serre and Patrick Tamain},
title = {Role of avalanche transport in competing drift wave and interchange turbulence},
journal = {Journal of Physics: Conference Series}
}

@article{krutkin2023method,
	title={A method for density fluctuation measurements using pulse reflectometry},
	author={Krutkin, Oleg and Brunner, Stephan and Coda, Stefano},
	journal={Nuclear Fusion},
	volume={63},
	number={7},
	pages={076012},
	year={2023},
	publisher={IOP Publishing}
}

@article{panico2025importance,
  title={On the importance of flux-driven turbulence regime to address tokamak plasma edge dynamics},
  author={Panico, Olivier and Sarazin, Y and Hennequin, P and G{\"u}rcan, {\"O}D and Bigu{\'e}, R and Dif-Pradalier, G and Garbet, X and Ghendrih, P and Varennes, R and Vermare, L},
  journal={Journal of Plasma Physics},
  volume={91},
  number={1},
  pages={E26},
  year={2025},
  publisher={Cambridge University Press}
}

@article{panico2025generation,
  title={Generation of zonal flows and impact on transport in competing drift waves and interchange turbulence},
  author={Panico, Olivier and Sarazin, Yanick and Hennequin, Pascale and G{\"u}rcan, Ozgur and Dif-Pradalier, Guilhe and Garbet, Xavier and Varennes, Robin},
  journal={Journal of Plasma Physics},
  volume={91},
  number={4},
  pages={E118},
  year={2025},
  publisher={Cambridge University Press}
}

@article{rienacker2025survey,
	title={Survey of the edge radial electric field in L-mode TCV plasmas using Doppler backscattering},
	author={Rien{\"a}cker, Sascha and Hennequin, Pascale and Vermare, Laure and Honor{\'e}, Cyrille and Coda, Stefano and Labit, Benoit and Vincent, Benjamin and Wang, Yinghan and Frassinetti, Lorenzo and Panico, Olivier and others},
	journal={Plasma Physics and Controlled Fusion},
	volume={67},
	number={6},
	pages={065003},
	year={2025},
	publisher={IOP Publishing}
}

@misc{pastore2026applicationsnovelmodelbasedrealtime,
      title={Applications of a novel model-based real-time observer for electron density profile control experiments in TCV}, 
      author={F. Pastore and O. Sauter and F. Felici and D. Kropackova and A. Balestri and C. Galperti and O. Kudlacek and K. Lee and A. Pau and T. Ravensbergen and S. Van Mulders and B. Vincent and N. M. T. Vu and the TCV team and the EUROfusion Tokamak Exploitation Team},
      year={2026},
      eprint={2603.26310},
      archivePrefix={arXiv},
      primaryClass={physics.plasm-ph},
      url={https://arxiv.org/abs/2603.26310}, 
}

@article{blanchard2019thomson,
  title={Thomson scattering measurements in the divertor region of the TCV tokamak plasmas},
  author={Blanchard, P and Andrebe, Y and Arnichand, H and Agnello, R and Antonioni, S and Couturier, S and Decker, J and DExaerde, T De Kerchove and Duval, BP and Furno, I and others},
  journal={Journal of Instrumentation},
  volume={14},
  number={10},
  pages={C10038--C10038},
  year={2019}
}

@PHDTHESIS{thesisBagnoto,
	url = "https://infoscience.epfl.ch/bitstreams/20c4cbdd-e4fb-4665-89d2-7806b1f589bb/download",
	title = "Study of impurity ion transport using charge exchange spectroscopy on TCV",
	author = "Bagnato, Filippo",
	school = "EPFL",
	year = "2022"
}

@article{PhysRevLett.33.758,
  title = {Electron Cyclotron Emission from a Tokamak Plasma: Experiment and Theory},
  author = {Costley, A. E. and Hastie, R. J. and Paul, J. W. M. and Chamberlain, J.},
  journal = {Phys. Rev. Lett.},
  volume = {33},
  issue = {13},
  pages = {758--761},
  numpages = {0},
  year = {1974},
  month = {Sep},
  publisher = {American Physical Society},
  doi = {10.1103/PhysRevLett.33.758},
  url = {https://link.aps.org/doi/10.1103/PhysRevLett.33.758}
}

@article{nazikian1995reflectometer,
  title={Reflectometer measurements of density fluctuations in tokamak plasmas},
  author={Nazikian, R and Mazzucato, E},
  journal={Review of scientific instruments},
  volume={66},
  number={1},
  pages={392--398},
  year={1995},
  publisher={American Institute of Physics}
}

@article{di2024system,
  title={System size scaling of triangularity effects on global temperature gradient-driven gyrokinetic simulations},
  author={Di Giannatale, Giovanni and Bottino, Alberto and Brunner, Stephan and Murugappan, Moahan and Villard, Laurent},
  journal={Plasma Physics and Controlled Fusion},
  volume={66},
  number={9},
  pages={095003},
  year={2024},
  publisher={IOP Publishing}
}

@article{hahm2018mesoscopic,
  title={Mesoscopic transport events and the breakdown of Fick’s law for turbulent fluxes},
  author={Hahm, TS and Diamond, PH},
  journal={Journal of the Korean Physical Society},
  volume={73},
  number={6},
  pages={747--792},
  year={2018},
  publisher={Springer}
}

@article{newman1996dynamics,
  title={The dynamics of marginality and self-organized criticality as a paradigm for turbulent transport},
  author={Newman, DE and Carreras, BA and Diamond, PH and Hahm, TS},
  journal={Physics of Plasmas},
  volume={3},
  number={5},
  pages={1858--1866},
  year={1996},
  publisher={American Institute of Physics}
}

@article{qi2020dimits,
  title={Dimits shift, avalanche-like bursts, and solitary propagating structures in the two-field flux-balanced Hasegawa--Wakatani model for plasma edge turbulence},
  author={Qi, Di and Majda, Andrew J and Cerfon, Antoine J},
  journal={Physics of Plasmas},
  volume={27},
  number={10},
  year={2020},
  publisher={AIP Publishing}
}

@article{conway1999reflectometer,
  title={A reflectometer for fluctuation and correlation studies on the Joint European Torus tokamak},
  author={Conway, GD and Vayakis, G and Fessey, JA and Bartlett, DV},
  journal={Review of scientific instruments},
  volume={70},
  number={10},
  pages={3921--3929},
  year={1999},
  publisher={American Institute of Physics}
}

@article{altukhov2016poloidal,
  title={Poloidal inhomogeneity of turbulence in the FT-2 tokamak by radial correlation Doppler reflectometry and gyrokinetic modelling},
  author={Altukhov, AB and Gurchenko, AD and Gusakov, EZ and Esipov, LA and Irzak, MA and Kantor, M Yu and Kouprienko, DV and Lashkul, SI and Leerink, S and Niskala, P and others},
  journal={Plasma Physics and Controlled Fusion},
  volume={58},
  number={10},
  pages={105004},
  year={2016},
  publisher={IOP Publishing}
}

@article{krutkin2019nonlinear,
  title={Nonlinear Doppler reflectometry power response. Analytical predictions and full-wave modelling},
  author={Krutkin, OL and Gusakov, EZ and Heuraux, St{\'e}phane and Lechte, C},
  journal={Plasma Physics and Controlled Fusion},
  volume={61},
  number={4},
  pages={045010},
  year={2019},
  publisher={IOP Publishing}
}

@phdthesis{panicothesis2024,
    url = "https://theses.fr/2024IPPAX133",
    author = "Panico, Olivier",
    title = "Edge turbulence self-organization in fusion plasmas",
    school = "Institut polytechnique de Paris",
    year = "2024"
}

@unpublished{gusakov2004analytical,
  TITLE = {{Analytical theory of Doppler reflectometry in slab plasma model}},
  AUTHOR = {Gusakov, Evgeniy and Surkov, Alexander},
  URL = {https://hal.science/hal-00001867},
  NOTE = {12th International Congress on Plasma Physics, 25-29 October 2004, Nice (France)},
  YEAR = {2004},
  MONTH = Oct,
  HAL_ID = {hal-00001867},
  HAL_VERSION = {v1},
}

@article{gusakov2004spatial,
  title={Spatial and wavenumber resolution of Doppler reflectometry},
  author={Gusakov, EZ and Surkov, AV},
  journal={Plasma physics and controlled fusion},
  volume={46},
  number={7},
  pages={1143--1162},
  year={2004}
}

@article{gusakov2005multiple,
  title={Multiple scattering effect in Doppler reflectometry},
  author={Gusakov, EZ and Surkov, AV and Popov, A Yu},
  journal={Plasma physics and controlled fusion},
  volume={47},
  number={7},
  pages={959--974},
  year={2005}
}

@article{gusakov2002non,
  title={Non-linear theory of fluctuation reflectometry},
  author={Gusakov, EZ and Popov, A Yu},
  journal={Plasma physics and controlled fusion},
  volume={44},
  number={11},
  pages={2327--2337},
  year={2002}
}

@article{krutkin2024validation,
  title={Validation of short-pulse reflectometry turbulence measurements with a synthetic diagnostic},
  author={Krutkin, Oleg and Kumar, Umesh and Mazzi, S and Brunner, S and Coda, S and Rien{\"a}cker, S and Van Rossem, M and TCV Team},
  journal={Nuclear Fusion},
  volume={64},
  number={2},
  pages={026010},
  year={2024},
  publisher={IOP Publishing}
}

@phdthesis{marini2017thesis,
    url = "https://infoscience.epfl.ch/handle/20.500.14299/142172",
    author = "Marini, Claudio",
    title = "Poloidal CX visible light plasma rotation diagnostics in TCV",
    school = "Ecole Polytechnique Federale de Lausane",
    year = "2017"
}

@article{jolliet2012plasma,
  title={Plasma size scaling of avalanche-like heat transport in tokamaks},
  author={Jolliet, S and Idomura, Y},
  journal={Nuclear Fusion},
  volume={52},
  number={2},
  pages={023026},
  year={2012}
}

@article{molina2019vband,
author = {Molina Cabrera, P. and Coda, S. and Porte, L. and Smolders, A. and TCV Team},
    title = {V-band nanosecond-scale pulse reflectometer diagnostic in the TCV tokamak},
    journal = {Review of Scientific Instruments},
    volume = {90},
    number = {12},
    pages = {123501},
    year = {2019},
    month = {12},
    issn = {0034-6748},
    doi = {10.1063/1.5094850},
    url = {https://doi.org/10.1063/1.5094850},
    eprint = {https://pubs.aip.org/aip/rsi/article-pdf/doi/10.1063/1.5094850/15702571/123501_1_online.pdf}
}

@article{gorler2011flux,
  title={Flux-and gradient-driven global gyrokinetic simulation of tokamak turbulence},
  author={G{\"o}rler, Tobias and Lapillonne, Xavier and Brunner, Stephan and Dannert, Tilman and Jenko, Frank and Aghdam, Sohrab Khosh and Marcus, Patrick and McMillan, Ben F and Merz, Florian and Sauter, Olivier and others},
  journal={Physics of Plasmas},
  volume={18},
  number={5},
  year={2011},
  publisher={AIP Publishing}
}

@inproceedings{freethy2019advances,
  title={Advances in turbulence measurements using new Correlation ECE and nT-phase diagnostics at ASDEX Upgrade},
  author={Freethy, Simon J and G{\"o}rler, Tobias and Creely, Alex J and Conway, Garrard D and Denk, Severin S and Happel, Tim and Henniquin, Pascale and Koenen, Christian and White, Anne E and ASDEX Upgrade Team},
  booktitle={EPJ Web of Conferences},
  volume={203},
  pages={03001},
  year={2019},
  organization={EDP Sciences}
}

@article{hornung2013turbulence,
  title={Turbulence correlation properties measured with ultrafast sweeping reflectometry on Tore Supra},
  author={Hornung, Gr{\'e}goire and Clairet, Fr{\'e}d{\'e}ric and Falchetto, GL and Sabot, Roland and Arnichand, Hugo and Vermare, Laure},
  journal={Plasma Physics and Controlled Fusion},
  volume={55},
  number={12},
  pages={125013},
  year={2013},
  publisher={IOP Publishing}
}

@article{guillon2026anisotropic,
  title={Anisotropic truncation for turbulent transport and zonal flows in the Hasegawa-Wakatani system},
  author={Guillon, Pierre and Angles, Robin and Sarazin, Yanick and Gurcan, Ozgur D},
  journal={Plasma Physics and Controlled Fusion},
  year={2026}
}

@article{theiler2026progress,
  title={Progress and innovations in the TCV tokamak research programme},
  author={Theiler, Christian and Adamek, J and Agostini, M and Albert, C and Alberti, S and Alberti, G and Aleiferis, S and Alessi, E and Anastassiou, G and Andr{\`e}be, Y and others},
  journal={Nuclear Fusion},
  volume={66},
  number={11},
  pages={116007},
  year={2026},
  publisher={IOP Publishing}
}

@article{garbet2004profile,
  title={Profile stiffness and global confinement},
  author={Garbet, X and Mantica, P and Ryter, F and Cordey, G and Imbeaux, F and Sozzi, C and Manini, A and Asp, E and Parail, V and Wolf, R},
  journal={Plasma physics and controlled fusion},
  volume={46},
  number={9},
  pages={1351--1373},
  year={2004}
}

@book{marple2019digital,
  title={Digital spectral analysis},
  author={Marple Jr, S Lawrence},
  year={2019},
  publisher={Courier Dover Publications}
}

@article{krutkin2020theoretical,
  title={A theoretical investigation of the turbulent structures tilting measurements with radial correlation Doppler reflectometry},
  author={Krutkin, OL and Gusakov, EZ and Heuraux, St{\'e}phane},
  journal={Plasma Physics and Controlled Fusion},
  volume={62},
  number={4},
  pages={045004},
  year={2020},
  publisher={IOP Publishing}
}

@article{krutkin2020investigation,
  title={Investigation of nonlinear effects in Doppler reflectometry using full-wave synthetic diagnostics},
  author={Krutkin, OL and Altukhov, AB and Gurchenko, AD and Gusakov, EZ and Heuraux, St{\'e}phane and Irzak, MA and Esipov, LA and Kiviniemi, TP and Lechte, C and Leerink, S and others},
  journal={Plasma Science and Technology},
  volume={22},
  number={6},
  pages={064001},
  year={2020},
  publisher={IOP Publishing}
}

\appendix
\newpage
\section{Summary of the performed discharges}
\label{appendix: summary of the performed discharges}
\begin{table}[h!]
	\centering
	\renewcommand{\arraystretch}{0.9}
	\captionsetup{width=.95\textwidth}
	\begin{tabular}{@{} >{\centering\arraybackslash}m{1.5cm} | >{\centering\arraybackslash}m{3cm} | >{\centering\arraybackslash}m{2cm} | >{\centering\arraybackslash}m{3cm} | >{\centering\arraybackslash}m{2.3cm} | >{\centering\arraybackslash}m{2.2cm} @{}}
		\toprule
		\textbf{Shot} & \textbf{Time window $[s]$} & \textbf{ECH} [kW] & \textbf{NBI-1} (ref) $[kW]$ & \textbf{SPR} \\ 
		\midrule 
		$82607$ & $[1.2 - 1.8]$ & $0$ & $140$  $\ $ ($170$) &  no \\
		$81069$ & $[1.4 - 1.8]$ & $0$ & $167$  $\ $ ($200$) & $81100$ \\
		$82607$ & $[0.6 - 1.2]$ & $0$ & $250$  $\ $ ($300$) & no \\
		$81084$ & $[0.6 - 1]$   & $0$ & $250$  $\ $ ($300$) & no \\
		$81065$ & $[0.6 - 1]$   & $0$ & $260$  $\ $ ($315$) & no \\
		$81084$ & $[1.4 - 1.8]$ & $0$ & $300$  $\ $ ($355$) & no \\
		$82615$ & $[1.2 - 1.8]$ & $0$ & $300$  $\ $ ($390$) & $82619$ \\
		$81065$ & $[1.4 - 1.8]$ & $0$ & $355$  $\ $ ($420$) & no \\
		$82615$ & $[0.6 - 1.2]$ & $0$ & $406$  $\ $ ($480$) & $82619$ \\
		$81069$ & $[0.6 - 1]$   & $0$ & $500$  $\ $ ($590$) & $81100$ \\
        $88942$ & $[1 - 1.8]$   & $0$ & $510$  $\ $ ($590$) & no \\
        $88943$ & $[1 - 1.8]$   & $0$ & $510$  $\ $ ($590$) & no \\
        $88944$ & $[1 - 1.8]$   & $0$ & $670$  $\ $ ($800$) & no \\
        $88945$ & $[1 - 1.8]$   & $0$ & $685$  $\ $ ($800$) & no \\
		\midrule
		$81087$ & $[0.6 - 1]$   & $590$ & $0$ & no \\
		$82611$ & $[0.6 - 1.2]$ & $590$ & $0$ & no \\
		$82611$ & $[1.2 - 1.8]$ & $810$ & $0$ & no \\
		$82612$ & $[0.6 - 1.2]$ & $890$ & $0$ & no \\
		$82612$ & $[1.2 - 1.8]$ & $1180$ & $0$ & no \\
		\bottomrule
	\end{tabular}
	\caption{Deuterium unfavourable USN shots for correlation measurements. Each row corresponds to a heating step plateau. ECH, and NBI-1 represent the total injected power. The reference value is also indicated for NBI-1 ($\sim 15 \%$ mismatch between set and injected values).}
	\label{table: performed deuterium tcv shot}
\end{table}

\section{Details on the MUSIC algorithm}
\label{appendix: details on the MUSIC algorithm}

In this appendix, we recall the principle of the MUSIC algorithm (detailed in ref \cite{marple2019digital}) and how it can be applied to DBS signals to extract the "instantaneous" Doppler frequency and more generally the signal dynamics. A more detailed explanation for the application to GAMs detection in DBS signals is given in ref.\cite{vermare2012detection})

\medskip

MUSIC is a frequency estimation algorithm in which the signal is modelled by a finite set of sinusoids (or complex exponentials) plus incoherent noise: 
\begin{align}
	x(j) =  \sum_{k=1}^{n_f} B_k e^{-i2\pi j f_k T} + n(j)
\end{align}
where $x(j)$ is the sampled signal (at time T, $N$ data points), $n_f$ is the number of frequency components, $B_k$, $f_k$ the amplitude and the frequency of the $k^{th}$ component and $n(j)$ the time dependent noise.
This method is particularly suited when there is an {\it a priori} rough knowledge of the signal frequency content, and a high frequency resolution is desired, even for short time series (small number of data points $N$) and low signal to noise ratio.

The central point, in this representation, is that the auto-correlation matrix $R = xx^\dagger$, formed from available complex data samples of $x(j)$,  can be decomposed into two orthogonal subspaces, the noise subspace spanned by the minimum eigenvector and the signal subspace spanned by the eigenvectors of interest (also assumed to be orthogonal). 

In this singular value decomposition, for a frequency corresponding to the signal frequency, the projection on the noise eigenvectors will take very small values ; a spectral estimator (pseudo-spectrum) can thus be formed by taking the inverse of this projection over the noise sub-space, which yields sharp peaks at the frequencies of the signal. In practice, the auto-correlation matrix (or covariance matrix) $R$ , is evaluated to approach statistically an ensemble averaged matrix , by evaluating a forward–backward averaged matrix formed from sampling the correlation matrix in smaller matrices of size $n_{corr} < N$. 

The key parameters for the decomposition are the number of frequencies $n_f$ in the signal (when the spectrum is composed by a unique Doppler component, this is reduced to one) and the size of the correlation matrix $n_{corr}$. It must be at least half the Doppler period to obtain an accurate eigenvalue evaluation. 

\medskip

\begin{figure}[h]
	\centering
	\includegraphics[width=1.05\textwidth]{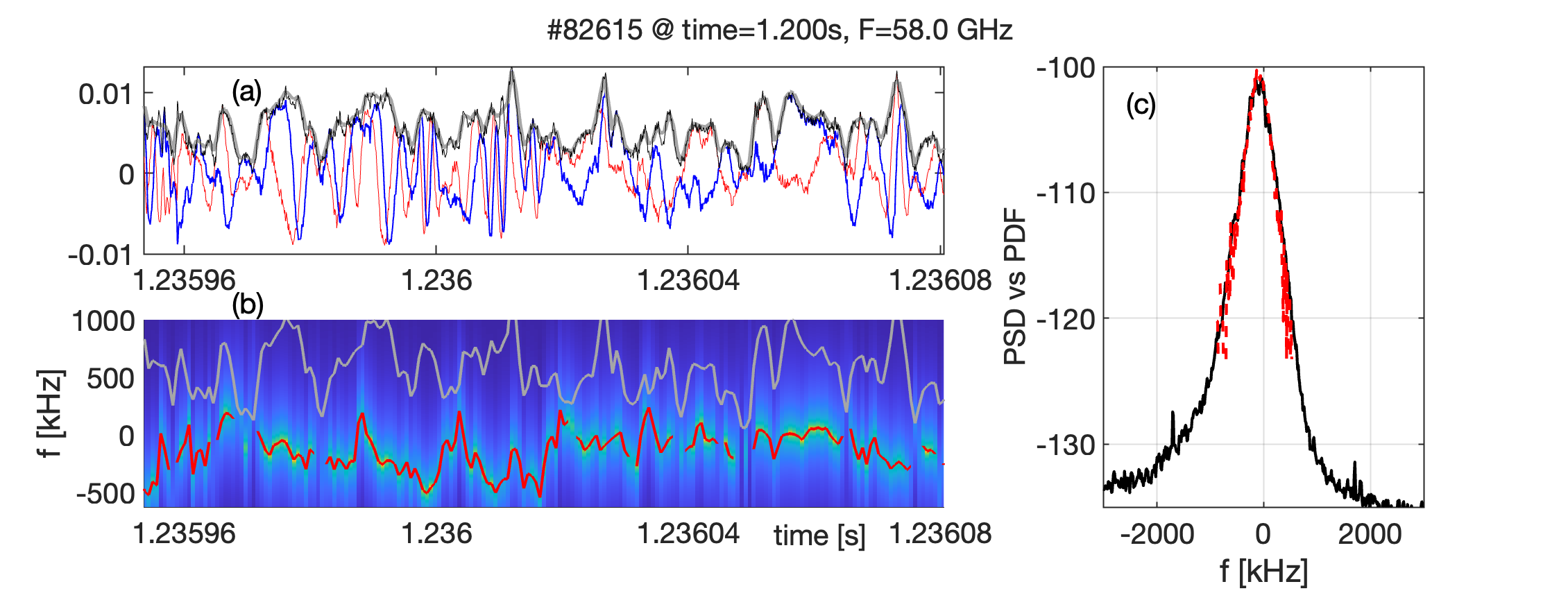}
	\caption{Example of DBS time series and their PSD for shot TCV $\#82615$, at probing frequencies $F=$\SI{58}{GHz} (upper panel) with a good signal to noise ratio, (a) Zoom (0.18 ms)  in the real (blue) and imaginary (red) part of the complex DBS signal, showing the successive fast or slow oscillations, signature of fast or slow fluctuations passing through the beam. The MUSIC 1st eigenvalue time series (grey bold line) captures the raw amplitude (black) dynamics. (b) Time-frequency analysis using the MUSIC algorithm: pseudo-spectrum time evolution, on which is superimposed the MUSIC eigenvalue (grey bold line). The frequency of the maximum of the pseudo spectrum is extracted in each small time window, providing the Doppler frequency time evolution (red). (c) Comparison of the power spectral density of the DBS raw signal and the probability distribution function of the Doppler instantaneous frequency from MUSIC (red).}
	\label{figapp: music good snr}
\end{figure}

This method can be applied to DBS signals to extract an "instantaneous" Doppler frequency and the corresponding amplitude time evolution. Indeed, the complex DBS signal provides instantaneous information on the velocity and amplitude of density fluctuations through the phase and amplitude of the signal, respectively. Standard spectral analysis of the signal instead provides the mean velocity, inferred from the mean Doppler shift. In practice, the latter is obtained from the Power Spectral Density (PSD), shown in \autoref{figapp: music good snr}(c), calculated from a few-millisecond DBS time series using the Welch algorithm, which reduces the variance of the spectral estimate. The average fluctuation amplitude is obtained from the integral of the PSD. The PSD consists of one single but broad Doppler peak, related to the statistical distribution of the fluctuations velocity. When examining a zoom in the time series, \autoref{figapp: music good snr} (a), the random signal appears as  successive bursts of fluctuations, the Doppler frequency (visible from the oscillation of the real/imaginary part of the signal -- red/blue) changing from burst to burst. The Doppler frequency is nearly constant in each of these small "burst" time intervals.

\medskip

\begin{figure}[h]
	\centering
	\includegraphics[width=1.05\textwidth]{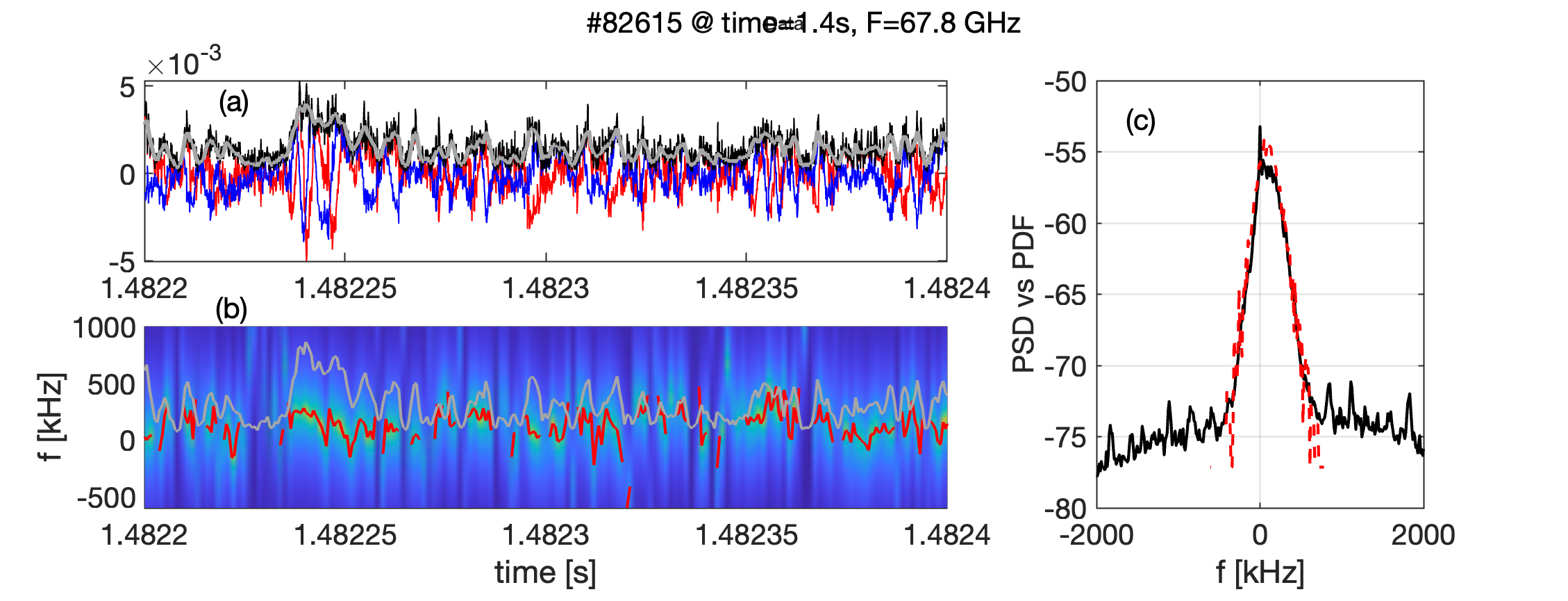}
	\caption{Example of DBS time series and their PSD for shot TCV $\#82615$, at probing frequencies $F=$\SI{67.8}{GHz} (upper panel) with a low signal to noise ratio, (a) Zoom (0.2 ms)  in the real (blue) and imaginary (red) part of the complex DBS signal, showing the fast or slow oscillations. In black and grey are indicated the raw and MUSIC-inferred amplitudes, respectively. (b) Time-frequency analysis using the MUSIC algorithm: pseudo-spectrum time evolution, on which is superimposed the MUSIC eigenvalue (grey bold line). (c) Comparison of the power spectral density of the DBS raw signal and the probability distribution function of the Doppler instantaneous frequency from MUSIC (red).}
	\label{figapp: music bad snr}
\end{figure}

Extracting the temporal evolution of the Doppler frequency will then consist of dividing the original time series into overlapping small sequences of typical length $n_{window}$ around or larger than the signal correlation time, and applying the MUSIC algorithm to each of them. The Doppler frequency time series will be formed from the peak frequency of each pseudo-spectrum, and the amplitude time series, from the dominant eigenvalue of the signal subspace. The time resolution of these time series is fixed by the window size and the overlapping parameter, usually taken as $n_{window}/2$. In \autoref{figapp: music good snr} (a), the MUSIC amplitude is compared to the raw signal amplitude, showing that the dynamics of the signal is preserved with a good time resolution. In addition, the comparison of the PSD to the Probability Distribution Function of the Doppler frequency time series shows a very good agreement, when the signal to noise ratio is high \autoref{figapp: music good snr} (c) as well as when it is much lower \autoref{figapp: music bad snr} (c).  

\begin{figure}[h]
	\centering
	\includegraphics[width=0.98\textwidth]{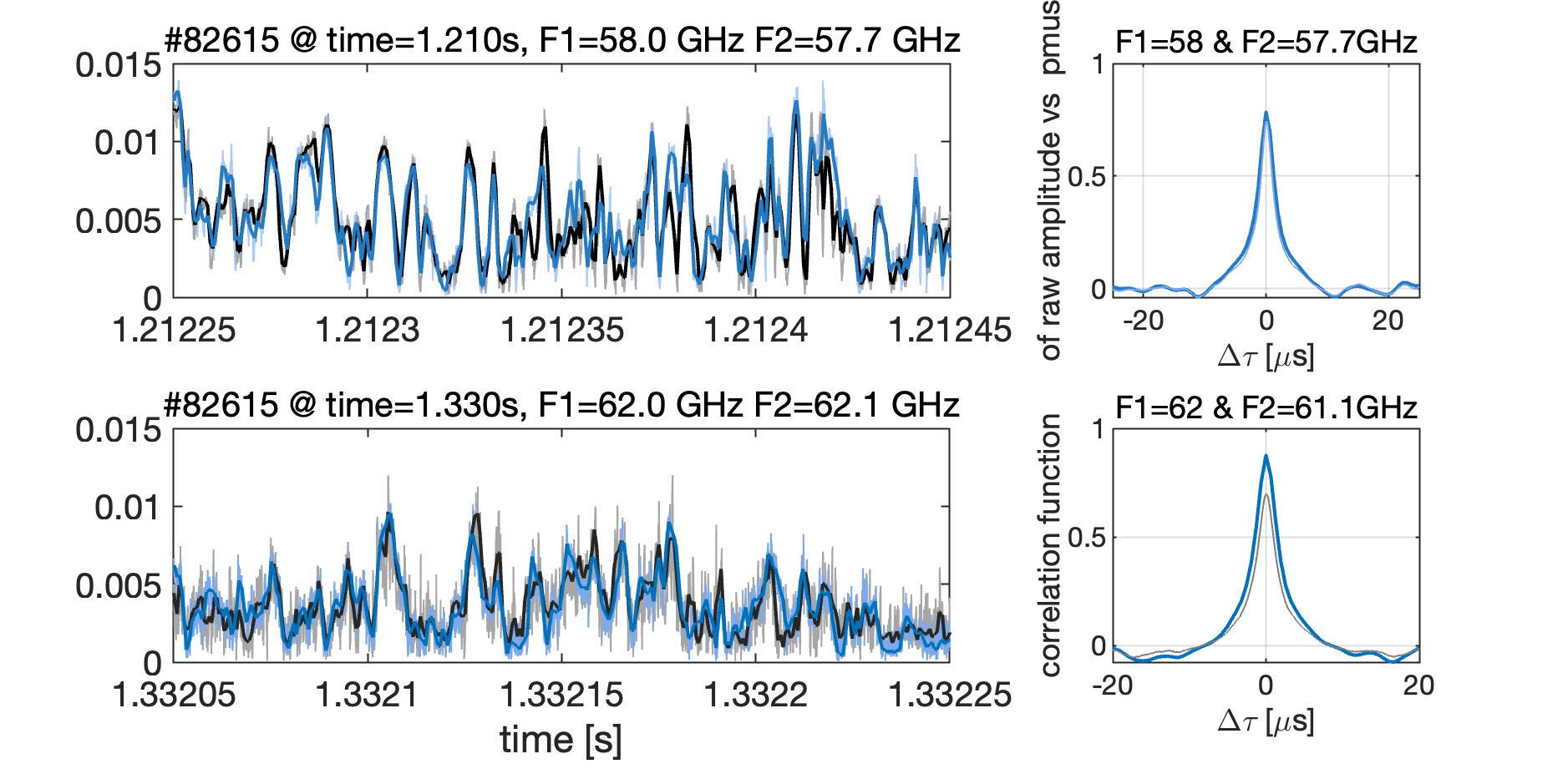}
	\caption{(a) MUSIC amplitude time series (bold line) superimposed to the raw amplitude (light line) for the two DBS signals probed at two close locations ($F=$\SI{58}{GHz} and $F=$\SI{57.7}{GHz}), in the case of good signal to noise ratio ; (c) in the case of low signal to noise ratio ($F=$\SI{62} and \SI{62.1}{GHz}). Comparison of the associated temporal correlation function obtained from the raw amplitude (light line) and from the MUSIC inferred amplitude (bold) shows a good agreement in the case of good signal to noise ratio (b) and improves the degraded correlation function in case of low signal to noise ratio (d). shot TCV $\#82615$.}
	\label{figapp: music corr}
\end{figure}

Taking advantage of its denoising capability, the MUSIC algorithm is applied to signals obtained at 2 different probing frequencies, (\autoref{figapp: music corr} left panels). In \autoref{figapp: music corr} (right panels) the correlation function computed from the MUSIC amplitude signals is compared to its direct evaluation. A very similar result is obtained in case of good signal to noise ratio (\autoref{figapp: music corr} (b). In case of low signal to noise ratio (\autoref{figapp: music corr} (d)) the level of correlation is improved and doesn't drop fast in the case of very close probing locations, where a still good correlation is expected.

\end{document}